\documentclass[%
 reprint,
 amsmath,amssymb,
 aps,
]{revtex4-2}

\usepackage{graphicx}
\usepackage{dcolumn}
\usepackage{bm}

\begin{document}

\preprint{APS/123-QED}

\title{First passage time and information of a one-dimensional Brownian particle with stochastic resetting to random positions.}

\author{J. Quetzalc\'oatl Toledo-Mar\'in }
 \email{j.toledo.mx@gmail.com}
\affiliation{%
 University of British Columbia, \\
  Vancouver BC, V6T 1Z4, Canada\\
  BC Children's Hospital Research Institute \\
  Vancouver BC, V6H 3N1 Canada 
}%


\author{Denis Boyer}
\affiliation{
 Instituto de F\'isica, \\
 Universidad Nacional Aut\'onoma de M\'exico\\
 C.P. 04510, CDMX, M\'exico \\
}%


\date{\today}

\begin{abstract}
We explore the effects of stochastic resetting to random positions of a Brownian particle on first passage times and Shannon's entropy. We explore the different entropy regimes, namely, the \textit{externally-driven}, the \textit{zero-entropy} and the \textit{Maxwell demon} regimes. We show that the mean first passage time (MPFT) minimum can be found in any of these regimes. We provide a novel analytical method to compute the MFPT, the mean first passage number of resets (MFPNR) and mean first passage entropy (MFPE) in the case where the Brownian particle resets to random positions sampled from a set of distributions known \textit{a priori}. We show the interplay between the reset position distribution's second moment and the reset rate, and the effect it has on the MFPT and MFPE.  We further propose a mechanism whereby the entropy per reset can be either in the Maxwell demon or the externally driven regime, yet the overall mean first passage entropy corresponds to the zero-entropy regime. Additionally, we find an overlap between the dynamic phase space and the entropy phase space. We use this method in a generalized version of the Evans-Majumdar model by assuming the reset position is random and sampled from a Gaussian distribution. We then consider the \textit{toggling reset} whereby the Brownian particle resets to a random position sampled from a distribution dependent on the reset parity. All our results are compared to and in agreement with numerical simulations. 
\end{abstract}

\maketitle


\section{Introduction}
Random processes with stochastic resetting exhibit diffusive and first passage features qualitatively different to ordinary diffusion \cite{evans2011diffusion, evans2011diffusionA, EvansJPhysA2013, ReuveniPRL2016, pal2016diffusion, pal2017first, toledo2019predator, gupta2019stochastic, evans2020stochastic}. When a Brownian particle in one dimension is stochastically reset to a fixed position in space, the mean first passage time to a given target becomes finite and can be minimized with respect to the resetting rate \cite{evans2011diffusion}. It has been shown that stochastic reset can enhance random searches in many contexts \cite{pal2017first, ReuveniPRL2016}, including situations where many targets or resetting points are distributed in space \cite{evans2011diffusionA}. Commonly, a resetting event is modelled as an instantaneous event where the Brownian particle relocates to a given position. While an instantaneous reset may not be physically possible, it still is a good approximation \cite{tal2020experimental}. In addition, non-instantaneous resetting processes have been studied \cite{maso2019transport}.

The statistics of first encounter times is at the core of many problems in the realm of biology, ecology, chemistry, economics, physics, etc. Good examples are reaction kinetics involving freely diffusive molecules and prey-predators dynamics \cite{krapivsky1996kinetics,redner1999capture, oshanin2009survival, gabel2012can, redner2014gradual}. In the latter context, prey capture can involve relatively complex decisions by the predator depending on the position of the prey, or vice-versa \cite{SchwarzlJPhysA2016,DasJPhysA2018}. For instance, this is the case when the predator uses information
about positions occupied by a prey, and decides to relocate to regions of space where it is more likely to be found \cite{mercado2018lotka}.
There are other phenomena, such as olfaction in the case of olfaction-driven navigation in animals 
\cite{boie2018information, baker2018algorithms, mercado2018lotka}, backtrack recovery in RNA polymerases \cite{dangkulwanich2013complete, roldan2016stochastic} or the formation of physical contacts 
between distant segments of DNA by means of temporal and spatial motion scales \cite{zhang2016first}, which can affect first passage properties.
Moreover, responding to environmental cues, \textit{i.e.}, acquiring, storing, processing information and acting upon it in biological systems inevitably comes at an energetic cost \cite{mehta2012energetic, wolpert2016free}. The energy cost will foremost depend on the situation being considered but ultimately we can presume a trade-off between energy optimization and time optimization. For instance, in \cite{toledo2019predator}, a prey-predator dynamics considers scenarios where the predator relocates to early positions visited by the prey, or, on the contrary, to recent positions. In both scenarios the predator captures the prey in a finite time, yet the predator's mean squared displacement yields a constant value in the former scenario whereas in the latter the predator diffuses like the prey. In addition, the regulatory functions, such as acquiring, storing and processing the prey information for relocation in the aforementioned example, need to be considered as part of the energetic cost.

The relationship between information and thermodynamics is a fascinating area of research which has led to a vast and still growing amount of research. Perhaps the main pillar linking information and thermodynamics is \textit{Landauer's principle} \cite{landauer1991information} which states that the irreversible erasure of information is accompanied with a dissipated energy in the form of heat. Some time ago, Landauer's principle was successfully verified experimentally \cite{blickle2006thermodynamics, berut2012experimental, jun2014high}. To illustrate Landauer's principle let's consider the irreversible erasure of information in one bit, whereby we initially have some information stored in this bit (either 0 or 1). Then by erasing we reset the bit to 0 regardless of the initial information stored, therefore losing the initial information stored. From Shannon's information theory \cite{mackay2003information}, the entropy associated to the initial state is $s_i = k \ln 2 $ while the entropy associated with the final state is $s_f=0$. Therefore the change in entropy is $\Delta s = -k \ln 2$. Landauer argued that by reducing the randomness of the physical entity (the bit in question) which must obey the laws of physics then the physical entity must manifest this change by dissipating an amount of energy equal to $kT \ln 2$.

In Ref. \cite{fuchs2016stochastic} the authors study the effects of Landauer's principle in stochastic resetting processes and identify three resetting regimes, namely, a regime where the particle requires external energy to reset referred to as the \textit{externally-driven regime}, the \textit{Maxwell Demon} regime whereby the particle is dissipating energy from the reservoir at each resetting and the \textit{zero-entropy} regime. Moreover, the authors derive the first and second law of thermodynamics for stochastic dynamics with resetting. In the present paper we study the first passage time and entropy properties in the presence of stochastic reset where Brownian particle relocates to a position sampled from a distribution. We also identify the three regimes mentioned above. Furthermore, we generalize a method discussed in previous work \cite{toledo2019predator} (a similar approach can also be found in \cite{ besga2020optimal}). This method allows a generalization where the reset position can be random variables from a known set of distributions, which we refer to as the \textit{resetting distribution set}. The resetting distribution set can, in principle, be a countable infinite set and dependent on the number of resets. We show the interplay between the resetting distribution's second moment and the reset rate, and the effect it has on the MFPT and MFPE.  We further propose a mechanism whereby the entropy per reset can be either in the \textit{Maxwell demon} or the \textit{externally-driven} regime, while the overall mean first passage entropy corresponds to the \textit{zero-entropy} regime. We argue these ideas can help guide the modeling of optimization processes akin to biology, condensed matter and machine learning among other fields. 

The paper is organized as follows: In the next section we derive the method to compute the mean first passage time (MFPT) and the mean first passage number of resets (MFPNR). In section \ref{sec:entropy} we outline the method and derive an analytical expression to compute the mean first passage entropy (MFPE). In section \ref{sec:evSat} we apply the method to a generalized version of the well-known Evans-Majumdar stochastic reset model \cite{evans2011diffusion} and compare the results with numerical simulations. We show that for the generalized Evans-Majumdar model, the MFPT minimum can be in any of the three regimes. Moreover, as the resetting distribution's second moment decreases the MFPT minimum goes from the externally-driven regime to the Maxwell demon regime passing through the zero-entropy regime. In section \ref{sec:TogGauss} we consider the \textit{toggling reset} whereby the Brownian particle resets to a random position sampled from a distribution dependent on the resetting event parity. We show that the entropy per reset can be in either the Maxwell demon or the externally-driven regime, yet the overall mean first passage entropy corresponds to zero-entropy regime. We further show the phase diagram where this net zero-entropy mechanism is possible. Section \ref{sec:Conclusions} is devoted to conclusions. The numerical simulations were done in Julia and the code is available in \cite{githubJaque}. More details on how the simulations were carried out can be found in \cite{toledo2019predator} and section \ref{sec:Simulations} of the SM.

\section{First Passage Dynamics under Stochastic reset to random positions}
\begin{figure}[hbtp]
\centering
\includegraphics[width=3.3in]{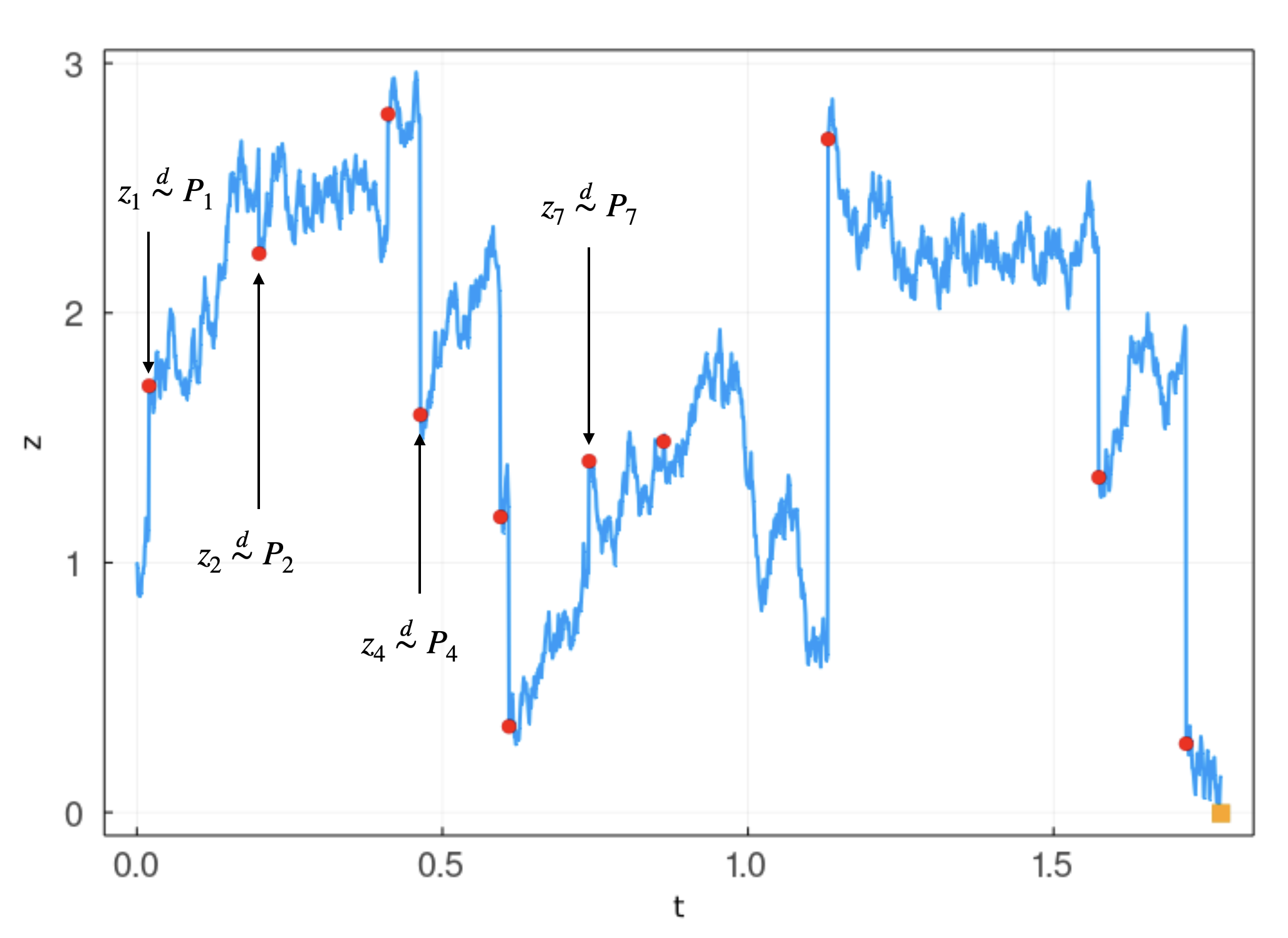}
\caption{Depiction of a Brownian particle under stochastic resetting (blue curve) with resetting rate $Q$. The resetting positions $\lbrace z_i \rbrace$ (red points) are independent random variables, where the resetting position $z_i$ is sampled from a distribution $P_i$ (we have pointed out some of the resetting positions). The set of distributions $\lbrace P_i \rbrace$ is the resetting distribution set. The process halts at the orange square, when the particle reaches the origin for the first time.} \label{fig:StochRes}
\end{figure}

Here we derive the formulas to compute the mean first passage time at the origin and the mean first passage number of resets for a one-dimensional Brownian particle under stochastic resets to random positions. Contrary to the problems studied in \cite{evans2011diffusion} or \cite{besga2020optimal}, the resetting position can be sampled at each resetting event from a different distribution. The Brownian particle is diffusing with a constant $D$, has an initial position $z_0$ distributed according to $P_0$ and stochastically resets with rate $Q$. The position $z_n$ of the $n$-th resetting event is drawn from a distribution $P_n$. Hence $n$ tags the number of resets and $\lbrace P_n \rbrace_{n=0}^{\infty}$ is a set of distributions, as shown in Fig. \ref{fig:StochRes}. We further show that under certain conditions the mean first passage time and the mean first passage number of resets are proportional with a proportionality constant $1/Q$. 
The formal mathematical problem can be expressed via a Fokker-Planck equation, namely:
\begin{eqnarray}
    \frac{\partial P(z,t)}{\partial t} &=& D\frac{\partial^2 P(z,t)}{\partial z^2} - QP(z,t) + Q P_n(z) \; \nonumber \\
    P(z,0) &=& P_0(z) \; , \label{eq:formalProblem}
\end{eqnarray}
where $P(z,t)$ denotes the particle distribution at position $z$ and time $t$. There are different methods for computing the mean first passage time and depending on the problem one method can adjust better than others. Here we use a method described in previous work \cite{toledo2019predator}. This method is motivated by \textit{i}) the fact that the mean first passage time relates to the survival probability in Laplace space \cite{zwanzig2001nonequilibrium} and \textit{ii}) each subprocess in-between resets is a simple Brownian motion. Given a resetting rate $Q$ and a set of resetting distributions $\lbrace P_i \rbrace_{i=0}^{\infty}$, we denote the survival probability as $S_Q(\lbrace P_i \rbrace_{i=0}^n, t)$, which is the probability that the particle has not reached the origin yet at time $t$.
We consider the initial positions and resetting positions to be random variables $\lbrace \xi_i \rbrace_{i=0}^{\infty}$ with distributions, $\lbrace P_{i} \rbrace_{i=0}^{\infty}$ known \textit{a priori}, \textit{i.e.}, $\xi_i \stackrel{\text{d}}{\sim} P_i$.
For a given sequence of probability density functions $\lbrace P_i \rbrace_{i=1}^n$, the probability that a process has survived and has reset exactly $n$ times at time $t$, $\mathcal{S}^{(n)}(\lbrace P_i \rbrace_{i=0}^n,t)$, may be expressed 
as the convolution of the survival 
probabilities of the diffusive process between two successive resetting events, $\langle \mathcal{S}(\xi_n,t_n) \rangle_{\xi_n}$, multiplied by the probability that a resetting event does not occur, where $t_n$ is the time interval between the $n$th and $n+1$th resetting events. Here $\langle \mathcal{S}(\xi_n,t_n) \rangle_{\xi_n}$ is averaged over $P_n(\xi_n)$. Notice that for $P_i(\xi_i) = \delta( \xi_i - z_i)$ the survival probability between any two consecutive resets is simply the survival probability at time $t_i$ of a Brownian particle with initial position $z_i$, \textit{viz.} $\mathcal{S}(z_i,t_i) = \text{erf}\left(|z_i|/\sqrt{4D t_i} \right)$, where $\text{erf}(\bullet)$ is the error function. 

From now on, we assume $Q \neq 0$ unless stated otherwise. First, let us derive the survival probability when there is only 1 reset.
The survival probability of the whole process is, in fact, the subprocess before the resetting event times the probability the reset does not occur convoluted with the subprocess after the resetting event times the probability the reset does not occur. Although we are considering only one reset, we include the probability the reset does not occur in both terms of the convolution as in the end we will consider the limit for infinite number of resets. The survival probability of a process with exactly 1 reset is
\begin{eqnarray}
\mathcal{S}^{(1)}(P_0,P_1, t)=Q\int_0^t dt_1 e^{-Qt_1} \langle \mathcal{S}(\xi_1,t_1) \rangle_{\xi_1}  \\
 \times  \int_0^t dt_0 e^{-Qt_0} \langle \mathcal{S}(\xi_0,t_0) \rangle_{\xi_0} \delta(t-t_1-t_0) \; . \nonumber
\end{eqnarray}
Generalizing to $n>0$ resets is straightforward, \textit{viz.},
\begin{eqnarray}
\mathcal{S}^{(n)}(\lbrace P_i \rbrace_{i=1}^n,t)=\int_0^t dt_1 \dots \int_0^t dt_n \prod_{i=1}^n e^{-Qt_i} Q \langle \mathcal{S}(\xi_i,t_i) \rangle_{\xi_i} \nonumber \\
\times  \int_0^t  dt_0 e^{-Qt_0} \langle \mathcal{S}(\xi_0,t_0) \rangle_{\xi_0} \delta \left(t-\sum_{i=0}^n t_i \right) \; . \nonumber \\ \label{eq:Sn} 
\end{eqnarray}
Applying the Laplace transform, which we denote with a $\widetilde{\bullet}$, Eq. \eqref{eq:Sn} yields
\begin{equation}
\widetilde{\mathcal{S}}^{(n)}(\lbrace P_i \rbrace_{i=0}^n,u)=\prod_{i=1}^n \left[ Q \langle \widetilde{\mathcal{S}}(\xi_i,u+Q) \rangle_{\xi_i} \right] \langle \widetilde{\mathcal{S}}(\xi_0,Q+u) \rangle_{\xi_0} \; . \label{eq:SurvLapnResets}
\end{equation}
The survival probability of the process is the sum over the number of resets $n$ of the survival probability with a fixed number of resets, \textit{viz.}
\begin{equation}
    S_Q(\lbrace P_i \rbrace_{i=0}^{\infty}, t) = \langle \mathcal{S}(\xi_0, t) \rangle e^{-Qt} + \sum_{n=1}^{\infty} \mathcal{S}^{n}(\lbrace P_i \rbrace_{i=1}^n, t) \; . \label{eq:SurvivalFormula}
\end{equation}
The survival probability functional (Eq. \eqref{eq:SurvivalFormula}) contains the first passage properties and can be used to extract the first passage time distribution. However, we will primarily focus on the mean first passage time (MFPT) functional.
Recall that the mean first passage time equates to the survival probability in Laplace domain in the limit where the Laplace variable, $u$, tends to zero.
Thus we apply Laplace transform to Eq. \eqref{eq:SurvivalFormula} leading to:
\begin{eqnarray}
\widetilde{\mathcal{S}}_Q(\lbrace P_i \rbrace_{i=0}^{\infty}  ,u )=\langle \widetilde{\mathcal{S}}(\xi_0,Q+u) \rangle_{\xi_0} + \sum_{n=1}^{\infty} \widetilde{\mathcal{S}}^{(n)}(\lbrace P_i \rbrace_{i=1}^n,u) \nonumber  \\
= \langle \widetilde{\mathcal{S}}(\xi_0,Q+u) \rangle_{\xi_0}\left(1+\sum_{n=1}^{\infty} \prod_{i=1}^n \left[ Q \langle \widetilde{\mathcal{S}}(\xi_i,u+Q) \rangle_{\xi_i} \right] \right) \; , \nonumber \\ \label{eq:S}
\end{eqnarray}
where we have used Eq. \eqref{eq:SurvLapnResets} in the last equality. 
We then obtain the MFPT, $\tau$, from the expression \eqref{eq:S},
\begin{eqnarray}
    \tau (Q, \lbrace P_i \rbrace_{i=0}^{\infty}) &=& \lim_{u\rightarrow 0 } \widetilde{\mathcal{S}}_Q(\lbrace P_i \rbrace_{i=0}^{\infty}  ,u ) \nonumber \\
    &=& \frac{\Theta(Q, P_0)}{Q} \left(1+\sum_{n=1}^{\infty} \prod_{i=1}^n \Theta(Q, P_i) \right) , \nonumber \\ \label{eq:tau}
\end{eqnarray}
where
\begin{eqnarray}
    \Theta(Q, P_n) &=& Q \langle \widetilde{\mathcal{S}}(\xi_n, Q) \rangle_{\xi_n} \nonumber \\
     &=& Q \int_{-\infty}^{\infty} d \xi_n P_n(\xi_n) \widetilde{\mathcal{S}}(\xi_n, Q) \; . \label{eq:Theta}
\end{eqnarray}
Notice that $\Theta(Q, \mathcal{P})$ is effectively the survival probability between two consecutive resets where the resetting position is averaged over a given distribution $\mathcal{P}$ and where the time interval between the two resetting events is averaged over the distribution $Qe^{-Qt}$. Consequently, $\Theta(Q, \mathcal{P})$ can also be interpreted as the probability of having survived when the next reset occurs. Similarly, $1 - \Theta(Q, \mathcal{P})$  is the probability of being absorbed before the next reset. Let us denote as $p_n(\lbrace P_i \rbrace_{i=0}^{n})$, the probability that the particle has survived after $n$ resets. From above, the probability of zero resets occurring is $p_0(Q, P_0) = 1 - \Theta(Q, P_0)$. Similarly, the event of only one reset correspond to the situation where the Brownian particle survives up to the first reset and does not survive between the first and the second ones. The probability of only one reset is thus $p_1(Q, P_0, P_1) = \Theta(Q,P_0) (1 - \Theta(Q, P_1))$. It is straightforward to show that the probability of finding the target placed at the origin after exactly $n$ resets, $p_n(\lbrace P_i \rbrace_{i=0}^{n})$, is:
\begin{equation}
    p_n(Q, \lbrace P_i \rbrace_{i=0}^{n}) = (1 - \Theta(Q, P_{n})) \prod_{i=0}^{n-1} \Theta(Q, P_i) \; . \label{eq:Pnresets}
\end{equation}
The product term in Eq. \eqref{eq:Pnresets} corresponds to the survival probability up to the $n$th reset while the prefactor accounts for the non-survival probability after the $n$th reset and before the $n+1$th reset. 

In the next section we define the mean first passage entropy. We show that the MFPE depends on two things, namely, the average change in entropy per reset and the mean first passage number of resets (MFPNR), defined as the average value of $n$ with the distribution \eqref{eq:Pnresets}:
\begin{equation}
    \mu(Q, \lbrace P_i \rbrace_{i=0}^{\infty}) = \sum_{n=0}^\infty n p_n(\lbrace P_i \rbrace_{i=0}^n) \; . \label{eq:mu}
\end{equation}
We leave the derivation of the average change in entropy per reset for the next section. 

Notice that when $\Theta(Q, P_n)$ is independent of $n$ (\textit{e.g.}, $P_n = \mathcal{P}$ for all $n=0,...,\infty$) and $\Theta(Q, \mathcal{P})<1$, Eq. \eqref{eq:mu} takes the form:
\begin{equation}
 \mu(Q, \mathcal{P}) = \Theta(Q, \mathcal{P})/(1-\Theta(Q, \mathcal{P})) \; .   \label{eq:MFPNRSim}
\end{equation}
whereas the MFPT in Eq. \eqref{eq:tau} simply reads:

\begin{equation}
    \tau (Q, \mathcal{P}) = \frac{\mu(Q, \mathcal{P})}{Q} \; . \label{eq:MFPTSim}
\end{equation}
From Eq. \eqref{eq:MFPNRSim}, the MFPNR is thus given by the ratio between the probability of at least one reset and the probability of no reset. The main results in this section can be summarized by Eqs. \eqref{eq:SurvivalFormula}, \eqref{eq:Theta}, \eqref{eq:tau} and \eqref{eq:mu}. In the case of identically distributed ${z_i}$, Eqs. \eqref{eq:MFPNRSim} and \eqref{eq:MFPTSim} are equivalent to the result derived in \cite{evans2018run} and \cite{evans2020stochastic} by using a renewal approach for the MFPT of an arbitrary process under resetting. These results relate the first passage statistics of the resetting process in terms of the ones corresponding to the process without resetting. In the following section we discuss the first passage properties under stochastic resetting from an information theory standpoint and derive the mean first passage entropy, \textit{i.e.}, the amount of entropy exchanged during the first passage process on average.

\section{First Passage entropy under Stochastic reset to random positions} \label{sec:entropy}
In this section we derive and discuss the change in entropy due to stochastic resetting to random positions. 
We show that there is an interplay between the resetting rate and the set of resetting distributions which yield three regimes, namely, the \textit{zero-entropy} regime where the entropy does not change on average until the first passage time, a regime where the entropy increases which we refer to as the \textit{externally-driven} regime and the \textit{Maxwell Demon} regime whereby the particle is dissipating or losing information every time it resets. We stress that these regimes refer to the effect of resetting on the information entropy, which contrasts with the monotonic increase in information entropy for simple diffusion due to the particle sampling of \textit{space}.

Let us consider the same model as in the previous section. At the $n-1$th resetting event, the particle relocates to a random position that we denote as $z_{n-1}$. Suppose that the next reset takes place $t$ units of time afterwards. The Brownian particle's entropy $s_{i}(t)$ before the $n$th reset is
\begin{eqnarray}
    s_{i}(t, z_{n-1}) &=& - \int_0^{\infty} dz \mathcal{N}_{im}(z | z_{n-1}, 2Dt) \nonumber \\ 
    && \times \ln \mathcal{N}_{im}(z | z_{n-1}, 2Dt) \; . \label{eq:s_i}
\end{eqnarray}
where $\mathcal{N}_{im}(z | z_n, 2Dt)$ denotes the normalized time-dependent distribution of a Brownian particle in the positive semi-axis with absorbing boundary conditions at the origin, \textit{i.e.}, $\mathcal{N}_{im}(z | z_n, 2Dt) = (\mathcal{N}(z | z_n, 2Dt) - \mathcal{N}(z | -z_n, 2Dt))/erf(z_n/\sqrt{4Dt})$. Here $\mathcal{N}(z | z_n, 2Dt)$ is a Gaussian distribution with mean and variance $z_n$ and $2Dt$, respectively.
The entropy after the $n$th reset, on the other hand, is related to the uncertainty one has about the position $z_n$:
\begin{equation}
    s(P_n) = - \int d \xi P_n(\xi) \ln P_n(\xi) \; .
\end{equation}
The entropy change $\Delta s(t, P_n)$ resulting from the $n$th resetting event is given by
\begin{eqnarray}
    \Delta s(t, P_n, z_{n-1}) &=& s(P_n) - s_{i}(t,z_{n-1}) \; . \label{eq:entropyChange}
\end{eqnarray}
We compute the entropy change per reset where the time $t$ between two consecutive resetting events is averaged over the distribution $Q\exp(-Qt)$ and averaged over the resetting distribution, namely,
\begin{equation}
    \Delta s_Q(P_n) = \int_0^{\infty} dt Q e^{-Qt} \langle \Delta s(t, P_n, z_{n-1}) \rangle_{\xi_{n-1}} \; . \label{eq:DeltaDifficult}
\end{equation}
The difficulty in computing Eq. \eqref{eq:DeltaDifficult} explicitly is due to Eq. \eqref{eq:s_i} and we are not going to attempt such calculation here. We consider a Gaussian distribution $\mathcal{N}(z|z_n, 2Dt)$ instead of $\mathcal{N}_{im}(z|z_n, 2Dt)$ in Eq. \eqref{eq:s_i}. In the following sections, we show that this approximation is in very good agreement with the numerical results at large resetting rate. When the typical resetting time is smaller than the typical diffusion time to the origin the Brownian particle does not have enough time to probe the boundary, which explains why at large resetting rate the analytical results yield very good agreement with the numerical ones. In the case of small resetting rates the analytical results show a decent quantitative agreement and the same qualitative behavior as obtained numerically. We argue that the qualitative agreement comes from the fact that the information entropy associated to, both, the Gaussian distribution and $\mathcal{N}_{im}(z | z_{n-1}, 2Dt)$ scales logarithmically with time.
The entropy change obtained from substituting the Gaussian distribution in Eq. \eqref{eq:s_i} yields

\begin{eqnarray}
    \Delta s(t, P_n) = - \int & d \xi_n & P_n(\xi_n) \ln P_n(\xi_n) \nonumber \\
    && - \frac{1}{2}(\ln 4\pi D t + 1) \; . \label{eq:deltaS}
\end{eqnarray}
Notice that Eq. \eqref{eq:deltaS} no longer has a dependence on $z_{n-1}$, hence we only average over the Poisson distribution, which yields

\begin{equation}
    \Delta s_Q(P_n) = s(P_n) + \frac{1}{2} \ln \frac{e^{\gamma - 1}Q}{4\pi D} \; . \label{eq:EntPerRes}
\end{equation}
where $\gamma \approx 0.57721...$ is the Euler's constant. Every time a reset occurs, there is an amount of entropy change which on average is given by Eq. \eqref{eq:EntPerRes}. We define the mean first passage entropy (MFPE), $\Delta \Sigma(Q, \lbrace P_n \rbrace_{n=0}^{\infty})$, as the sum over the number of resets of the entropy change per reset (Eq. \eqref{eq:EntPerRes}) averaged by the resetting probability, \textit{viz.}
\begin{equation}
    \Delta \Sigma(Q, \lbrace P_n \rbrace_{n=0}^{\infty}) = \sum_{n=1}^{\infty} \sum_{j=1}^n \Delta s_Q(P_j) p_n(\lbrace P_i \rbrace_{i=0}^n) \; . \label{eq:Entropy}
\end{equation}
Notice that when the resetting distributions are independent of the number of resets, $P_n = \mathcal{P}$ for all $n=0,...,\infty$, the MFPE is simply the mean entropy change per reset times the average number of resets, \textit{i.e.}, Eq. \eqref{eq:Entropy} can be simplified to
\begin{equation}
    \Delta \Sigma(Q, \mathcal{P}) = \Delta s_Q(\mathcal{P}) \mu(Q, \mathcal{P}) \; . \label{eq:EntropySim}
\end{equation}

Using Eq. \eqref{eq:Entropy} we obtain three different regimes, namely
\begin{equation}
    \begin{cases}
        \text{\textit{Externally-driven} regime}, \quad \Delta \Sigma(Q, \lbrace P_n \rbrace_{n=0}^{\infty}) >0 \\
       \text{\textit{Zero-entropy} regime} , \quad \Delta \Sigma(Q, \lbrace P_n \rbrace_{n=0}^{\infty}) =0 \\
        \text{\textit{Maxwell demon} regime} , \quad \Delta \Sigma(Q, \lbrace P_n \rbrace_{n=0}^{\infty}) <0 
    \end{cases} \label{eq:entropyRegimes}
\end{equation}
The regime where the average entropy is positive corresponds to the situation where information increases. In the \textit{Maxwell's demon} regime the system is losing information. However, Eq. \eqref{eq:entropyRegimes} allows search processes with stochastic resetting that generate zero entropy, i.e., no information is loss on average. Another interesting aspect of the previous derivations is the possibility of having a negative entropy change per reset for a subset of resetting events and still be in the \textit{zero-entropy} regime as long as there is enough positive entropy change per reset to compensate. We will come back to this later. The interplay between the set of resetting distributions and the resetting rate ultimately determine the entropy regime. The same interplay affects the mean first passage time and, as we will show in the next section, what optimizes the MFPT does not always optimize the MFPE.

In the next section we consider several of examples and compare our theoretical predictions with simulations. We show the existence of a metastable minimum at low resetting rates which can be tuned to fall in either of the entropy regimes. We center our focus on the MFPT metastable minimum mainly because i) it corresponds to a finite resetting rate compared to the global minimum which occurs for $Q\rightarrow \infty$, ii) it can be tuned to lie in any of the three regimes as per the global minimum belongs to the \textit{externally-driven} regime and iii) in the limit whereby the second moment of the resetting distribution tends to zero, the metastable minimum becomes the stable (and global) one.

\section{Examples}

\subsection{Stochastic resetting to random positions sampled from a Gaussian distribution} \label{sec:evSat}
The \textit{Evans-Majumdar} resetting model consists in a one-dimensional Brownian particle diffusing with a constant $D$ initially located at $z_0$, \textit{i.e.}, $P_0(\xi_0) = \delta(\xi_0-z_0)$ and stochastically resets to $z_0$ with a rate $Q$ \cite{evans2011diffusion}. A slightly more general model \cite{besga2020optimal} consists in assuming that both the initial condition and the resetting position are Gaussianly distributed, according to $\mathcal{N}(z|z_0,\sigma)$, with mean and standard deviation $z_0$ and $\sigma$, respectively, \textit{i.e.}, $P_n(\xi_n) = \mathcal{N}(\xi|z_0,\sigma)$ for all $n=0,...,\infty$. We are interested in the first passage properties derived in the previous sections, specifically, \textit{What is the mean time, number of resets and entropy of the Brownian particle when it reaches $z=0$ for the first time?}. We stress that all of the results related to the MFPT in this section are in perfect agreement with those obtained in \cite{besga2020optimal}. It's well-known that the Fokker-Planck that describes the process yields:
\begin{eqnarray}
    \frac{\partial P(z,t)}{\partial t} &=& D\frac{\partial^2 P(z,t)}{\partial z^2} - QP(z,t) + Q\mathcal{N}(z|z_0,\sigma) \; . \label{eq:hallmark} \nonumber \\
    P(z,0) &=& \mathcal{N}(z|z_0,\sigma) \label{eq:eq:FP-Gauss}
\end{eqnarray}

Indeed by making $\sigma \rightarrow 0$ in Eq. \eqref{eq:eq:FP-Gauss}, one recovers the Evans-Majumdar stochastic resetting model \cite{evans2011diffusion}.

In this example the MFPT, the MFPNR and the MFPE take very simple forms given by Eqs. \eqref{eq:MFPTSim}, \eqref{eq:MFPNRSim} and \eqref{eq:EntropySim}, respectively. The survival probability between two consecutive resets yields
\begin{eqnarray}
    \Theta(Q, \mathcal{N}) &=& 1 - \frac{1}{2} \left( e^{-|z_0| \sqrt{\frac{Q}{D}} + \frac{Q \sigma^2}{2D} } \text{erfc} \left( \frac{-|z_0| + \sigma^2 \sqrt{\frac{Q}{D}} }{\sqrt{2\sigma^2}} \right) \right. \nonumber \\
    && \left.+ e^{|z_0| \sqrt{\frac{Q}{D}} + \frac{Q \sigma^2}{2D}  } \text{erfc} \left( \frac{|z_0| + \sigma^2 \sqrt{\frac{Q}{D}} }{\sqrt{2\sigma^2}} \right) \right)\; . \label{eq:hallmarMFPT-2}
\end{eqnarray}
Note that the first passage properties are functions of $\Theta(Q,\mathcal{N})$. We will now discuss various limits.
By taking the limit $\sigma \rightarrow 0$, Eq. \eqref{eq:hallmarMFPT-2} becomes
\begin{equation}
    \lim_{\sigma \rightarrow 0} \Theta(Q, \mathcal{N}) = 1- e^{-\sqrt{\frac{Q}{D}}|z_0|} \; . \label{eq:previous}
\end{equation}
From Eqs. \eqref{eq:MFPTSim} and \eqref{eq:MFPNRSim} together with \eqref{eq:previous} we obtain the MFPT and MFPNR when the resetting distribution is a $\delta$-Dirac function,
\begin{equation}
    \begin{cases}
        \lim_{\sigma \rightarrow 0} \mu(Q, \mathcal{N}) = e^{\sqrt{\frac{Q}{D}}|z_0|} - 1 \; , \\
        \lim_{\sigma \rightarrow 0} \tau(Q, \mathcal{N}) = (e^{\sqrt{\frac{Q}{D}}|z_0|} - 1)/Q
    \end{cases} \; ,
\end{equation}
in perfect agreement with \cite{evans2011diffusion}.
The entropy change per reset averaged over the resetting distribution at finite $\sigma$, Eq. \eqref{eq:EntPerRes}, yields
\begin{equation}
    \Delta s_Q(\mathcal{N}) = \frac{1}{2} \ln(\frac{\sigma^2 Q e^\gamma}{2D}) \; , \label{eq:entoForGauss}
\end{equation}
which diverges when $\sigma \rightarrow 0$. Hence the MFPE (Eq. \eqref{eq:Entropy}) diverges to minus infinity when the variance goes to zero. This regime corresponds to the \textit{Maxwell demon} (see Eq. \eqref{eq:entropyRegimes}). Moreover, from Eq. \eqref{eq:entoForGauss} the ratio between the variance and the typical mean-squared displacement dictates the entropy regime. This ratio has also an effect on the MFPT and MFPNR. Let us introduce the adimensional numbers $c=|z_0|/\sqrt{2\sigma^2}$ and $w=\sqrt{Q\sigma^2/2D}$. Eq. \eqref{eq:hallmarMFPT-2} becomes:
\begin{eqnarray}
    \Theta(Q, \mathcal{N}) = 1 - \frac{1}{2} e^{-c^2}\left(e^{(w-c)^2}\text{erfc}(w-c) \right . \nonumber \\
    \left. + e^{(w+c)^2}\text{erfc}(w+c) \right) \label{eq:hallmarMFPT-3} \; .
\end{eqnarray}
The MFPT is obtained by substituting Eq. \eqref{eq:hallmarMFPT-3} into Eq. \eqref{eq:MFPTSim}, namely,
\begin{eqnarray}
     \tau &=& \frac{1 - \frac{1}{2} e^{-c^2}\left(e^{(w-c)^2}\text{erfc}(w-c) + e^{(w+c)^2}\text{erfc}(w+c) \right)}{e^{-c^2}\left(e^{(w-c)^2}\text{erfc}(w-c) + e^{(w+c)^2}\text{erfc}(w+c) \right)} \nonumber \\
    && \times \frac{z_0^2}{4c^2w^2 D} \; . \label{eq:MFPT_Gauss}
\end{eqnarray}
The MFPNR is obtained by substituting Eq. \eqref{eq:hallmarMFPT-3} into Eq. \eqref{eq:MFPNRSim}, while the MFPE is
\begin{equation}
    \Delta \Sigma = \frac{1}{2}\ln \left( w^2 e^{\gamma} \right) \; .
\end{equation}
The entropy regime limits shown in Eq. \eqref{eq:entropyRegimes} become
\begin{equation}
    \begin{cases}
        \text{\textit{Externally-driven} regime,} & w>\exp(-\gamma/2) \\
        \text{\textit{Zero-entropy} regime,} & w=\exp(-\gamma/2) \\
        \text{\textit{Maxwell demon} regime,} & w<\exp(-\gamma/2) \\
    \end{cases}
\end{equation}

To obtain the behavior of the MFPT for small and large resetting rates we expand Eq. \eqref{eq:MFPT_Gauss} in series of $w$  around $0$ and $\infty$, respectively, leading to:
\begin{equation}
    \tau \approx 
    \begin{cases}
        \frac{e^{-c^2}}{\sqrt{\pi} c^2 w} \frac{z_0^2}{2D}, \; w \ll 1 \\
        \frac{\sqrt{\pi} e^{c^2}}{2c^2 w} \frac{z_0^2}{2D}, \; w \gg 1
    \end{cases} \; . \label{eq:MFPT_limits}
\end{equation}

\begin{figure}[hbtp]
\centering
\includegraphics[width=3.3in]{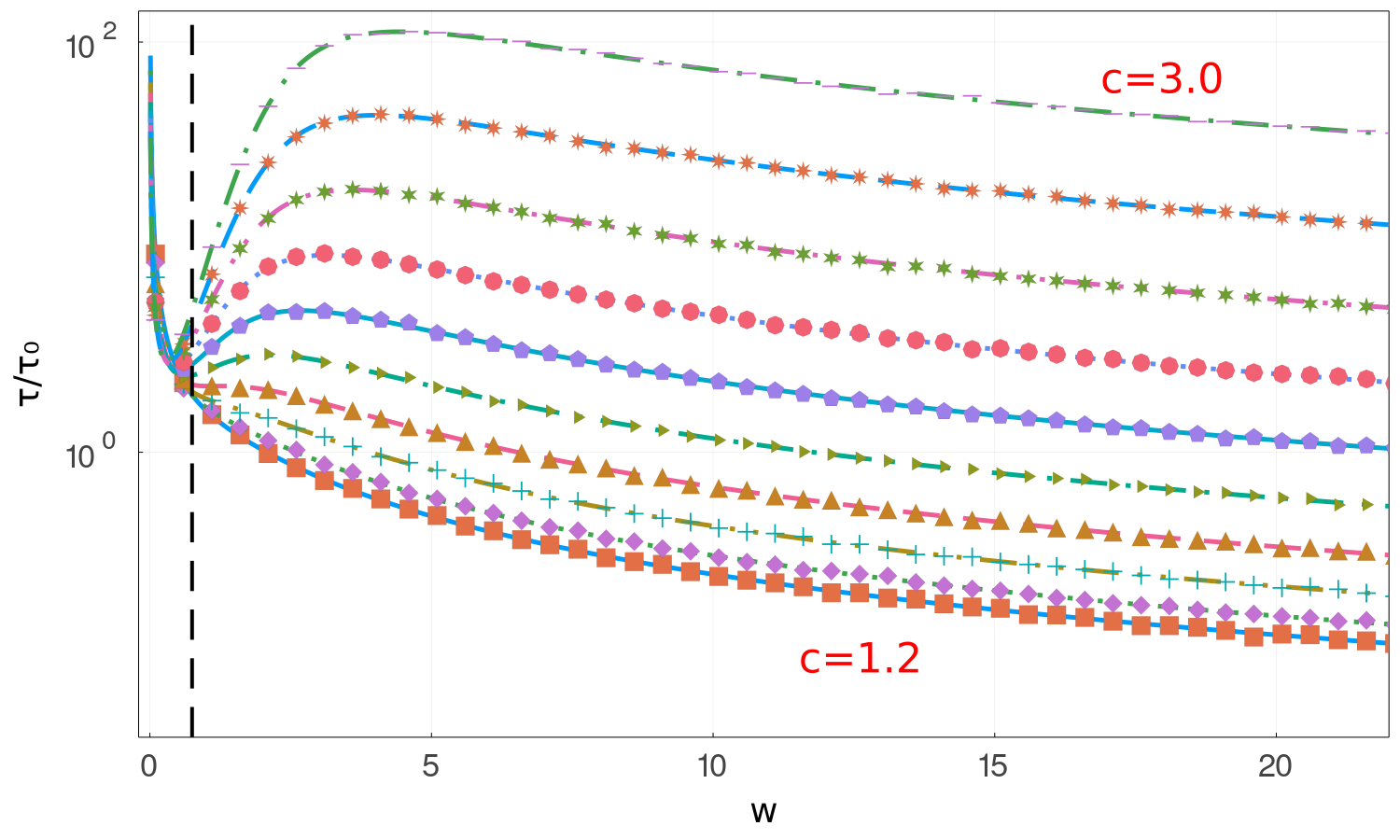}
\label{fig:FPTGauss}
\includegraphics[width=3.3in]{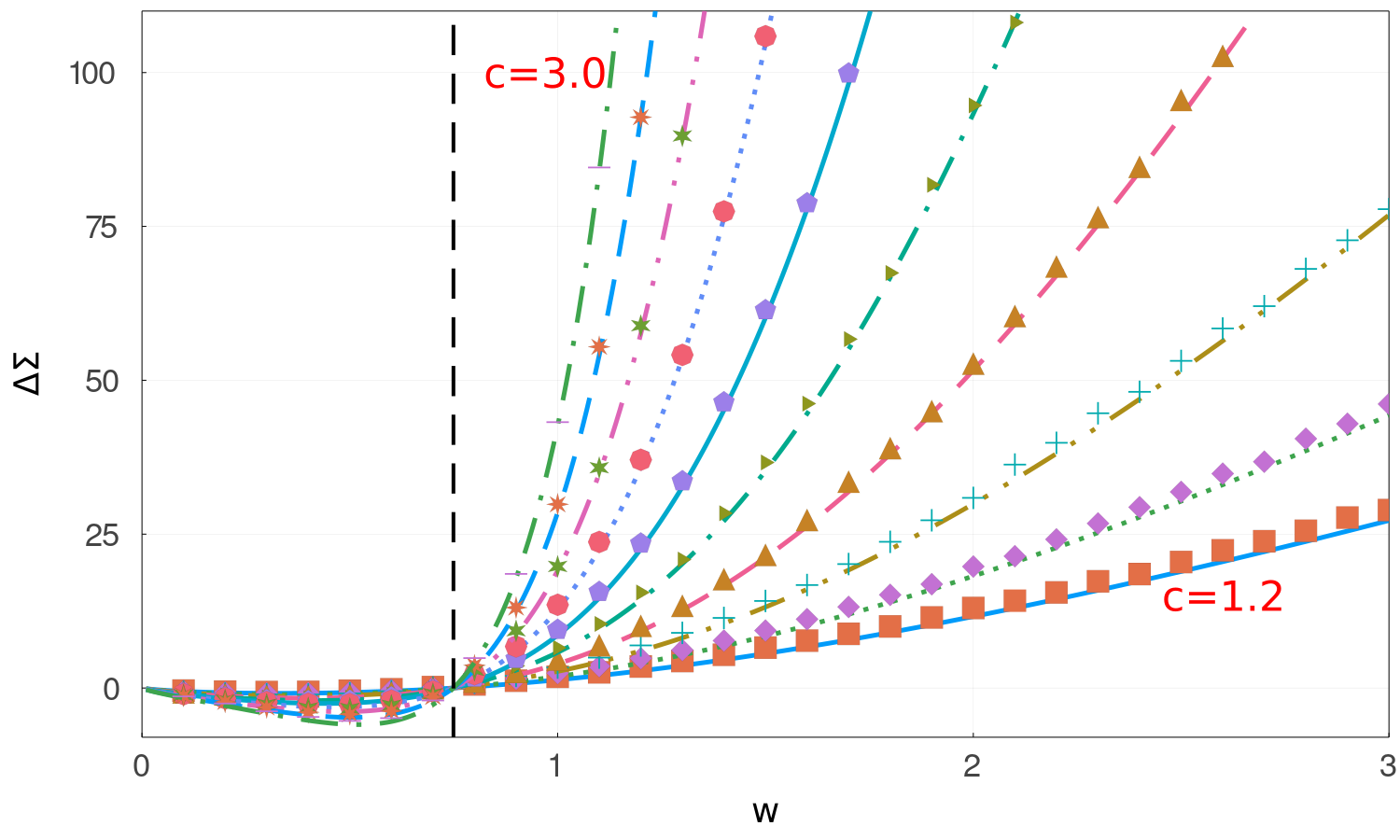}
\label{fig:MFPEC}
\caption{Comparison between theory and simulations for $c \in [1.2,3.0]$ taking steps of size $\Delta c = 0.2$. Lines and plot markers correspond to theory and simulations respectively. The dashed vertical line delimits the entropy regimes, \textit{i.e.}, left and right regions to the dashed line correspond to \textit{Maxwell demon} and \textit{externally-driven} resetting regimes, respectively, while at the dashed corresponds to the \textit{zero-entropy} regime. \textbf{Upper panel} The reduced MFPT \textit{vs} $w$, where $\tau_0 = z_0^2/2D$. For $c< 1.8$ the MFPT is decreasing monotonically. For $c\geq 1.8$, the MFPT has a metastable minimum close to the \textit{zero-entropy} regime. As $c$ increases to infinity, the metastable minimum becomes a stable one. \textbf{Lower panel} MFPE \textit{vs} $w$. All curves pass through the origin and the coordinates $(\exp(-\gamma/2), 0)$ corresponding to no reset and the \textit{zero-entropy} regime respectively. As $c$ increases, the MFPE deviates more from zero for all $w$ values but $0$ and $\exp(-\gamma/2)$. The entropy numerical computation is detailed in section \ref{sec:Simulations} of the SM. } \label{fig:Gauss}
\end{figure}

The MFPT for small and large resetting rates goes as $\tau \sim Q^{-1/2}$. While the MFPT diverges as $Q$ tends to zero as expected, for large resetting rate the MFPT tends to zero as a power law of $Q$. These behaviours are qualitatively different from the Evans-Majumdar resetting model \cite{evans2011diffusion}. In the original model, for large values of Q the particle has not enough time to reach the origin between resets. Here, a large resetting rate also increases the sampling of the resetting distribution such that eventually the particle resets to a position very close to the target. Perhaps more striking is the behavior for finite values of $Q$. For values of $c\geq 1.8$ the MFPT has a metastable minimum. In agreement with the results of \cite{besga2020optimal} where the authors showed that below a critical value $c\approx 1.79$ the metastable minimum disappears both theoretically and experimentally.

\begin{figure}[hbtp]
\centering
\includegraphics[width=2.5in, height=2.6in]{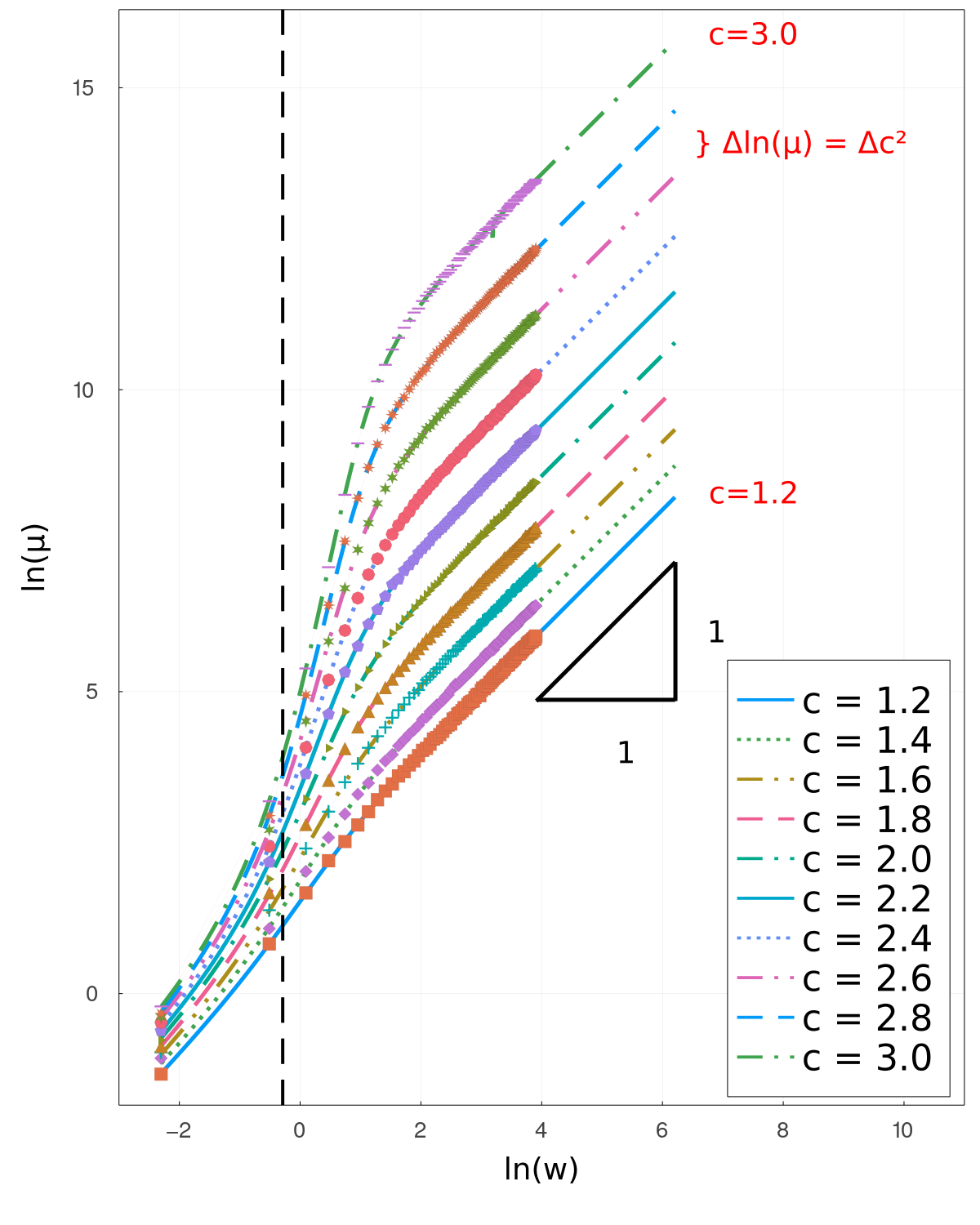}
\label{fig:MFPNR}
\includegraphics[width=2.9in]{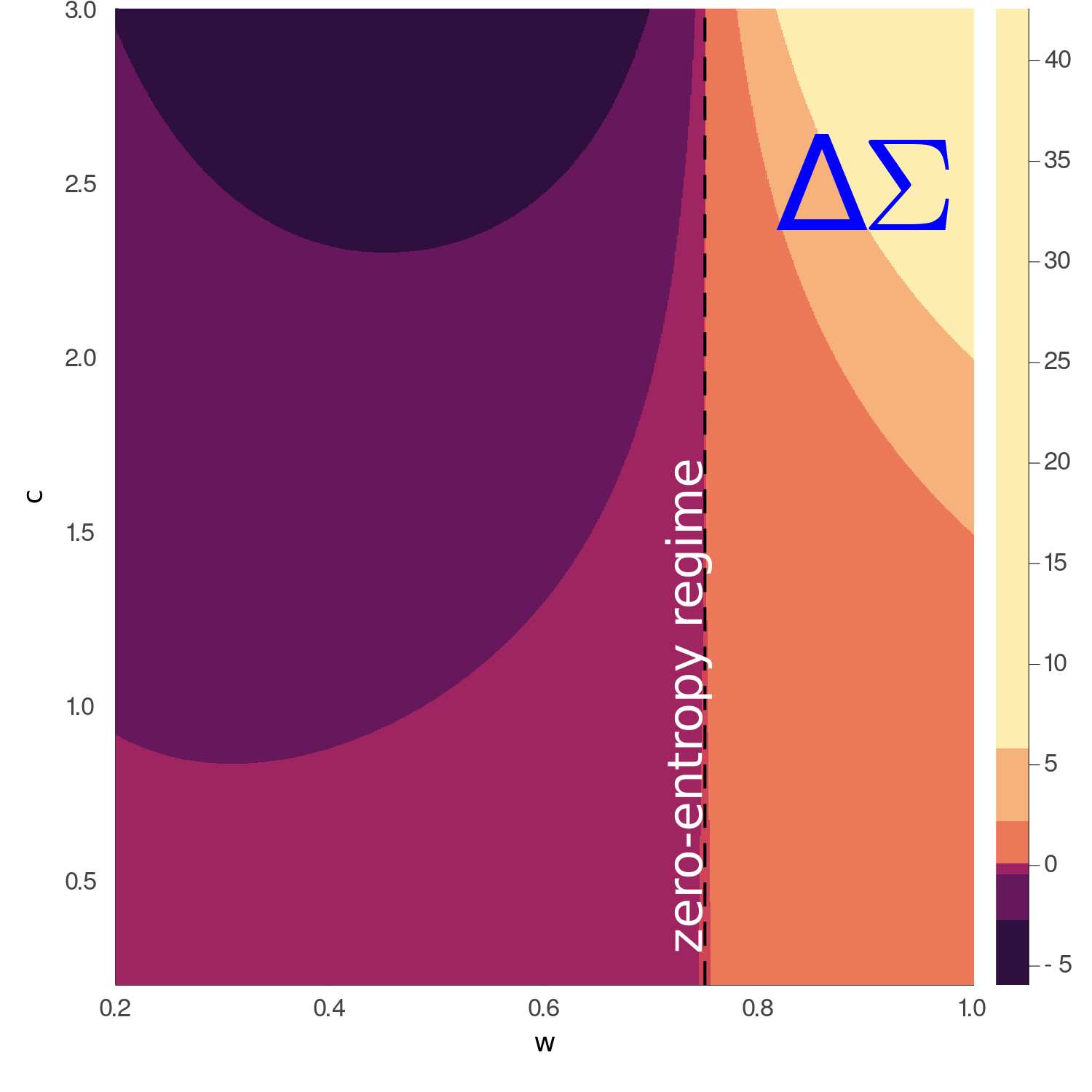}
\label{fig:MFPEC_cont}
\caption{Comparison between theory and simulations for values of $c \in [1.2,3.0]$ taking steps of size $\Delta c = 0.2$. Lines and plot markers correspond to theory and simulations respectively. The dashed vertical line delimits the entropy regimes, \textit{i.e.}, left and right regions to the dashed line correspond to \textit{Maxwell demon} and \textit{externally-driven} resetting regimes, respectively, while at the dashed corresponds to the , \textit{zero-entropy} regime. \textbf{Upper panel} Natural log of MFPNR \textit{vs} $\ln w$. For large values of $w$, $\ln \mu \sim c^2 + \ln w$ in agreement with Eq. \eqref{eq:MFPT_limits}. \textbf{Lower panel} Contour plot of the MFPE. In the \textit{Maxwell demon} regime, the magnitude of the deviations from zero are smaller than in the \textit{energy-driven} regime.} \label{fig:Gauss2}
\end{figure}

In the upper panel of Fig. \ref{fig:Gauss} we show the reduced MFPT $\tau /\tau_0$ \textit{vs} $w$, where $\tau_0=z_0^2/2D$, and compare theory (Eq. \eqref{eq:MFPT_Gauss}) with numerical simulations for different values of $c$. We have also plotted a vertical dashed line which delimits the entropy regimes, \textit{viz.}, $w$ values below, at and above the dashed line correspond to \textit{Maxwell demon}, \textit{zero-entropy} and \textit{externally-driven} resetting regimes, respectively. Notice that the MFPT metastability is located close to the \textit{zero-entropy} regime. As $c$ increases to infinity, the metastable minimum becomes a stable minimum, as for a single resetting position. In the lower panel of Fig. \ref{fig:Gauss2} we show the entropy change defined by $\Delta \Sigma$ \textit{vs} $w$ for different values of $c$. Notice that all curves pass through the point with coordinates $(\exp(-\gamma/2), 0)$. As $c$ increases, the MFPE deviates more from zero for all $w$ values except for $w=0$ and $w=\exp(-\gamma/2)$, which correspond to $Q=0$ and the \textit{zero-entropy} regime, respectively. In the upper panel of Fig \ref{fig:Gauss2} we show the MFPR $\mu$ \textit{vs} $w$ for different values of $c$ to validate our analytical results by comparing with simulations. Notice that for large values of $w$, $\ln \mu \sim c^2 + \ln w$ in agreement with Eq. \eqref{eq:MFPT_limits}. In the lower panel of Fig. \ref{fig:Gauss2} we show a contour plot of the MFPE. The vertical dashed line delimits the entropy regimes. 

\begin{figure}[hbtp]
\centering
\includegraphics[width=3.3in]{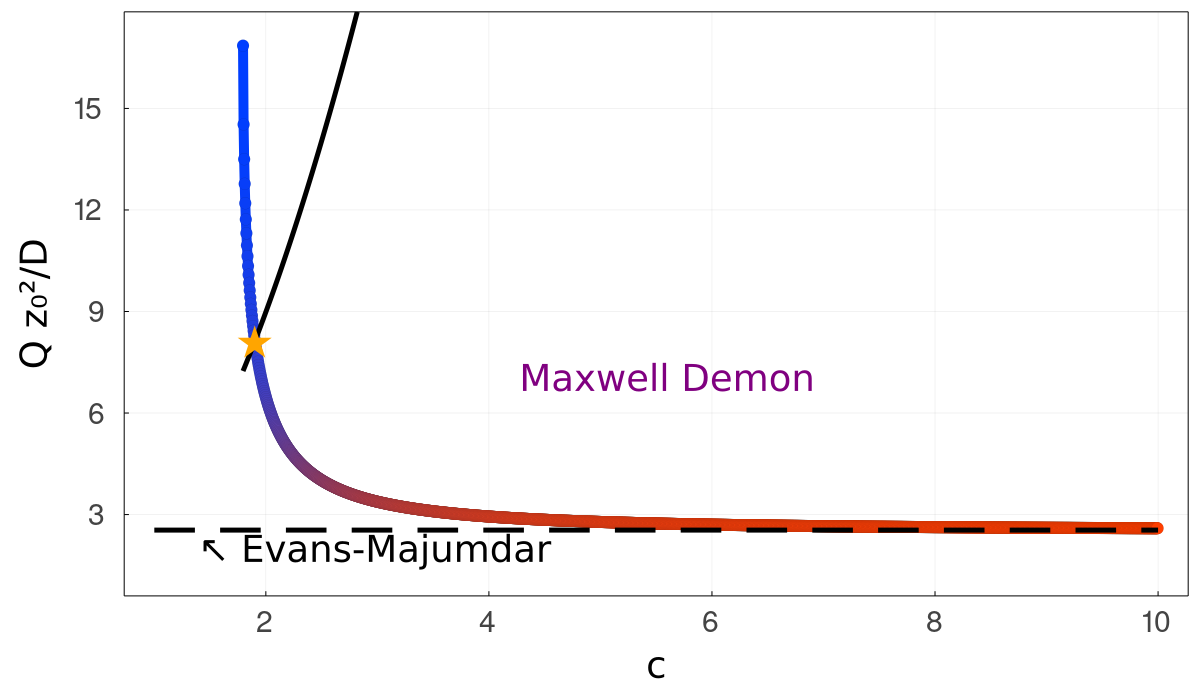}
\caption{Rescaled resetting rate which locally minimizes the MFPT \textit{vs} $c$. The local minima were obtained by numerically solving Eq. \eqref{eq:trans}. The data points are the numerical roots and the coloring depicts the MFPT value, where blue corresponds to a lower MFPT whereas red corresponds to higher values. The horizontal black dashed line corresponds to the reduced resetting rate that minimizes the standard model of \cite{evans2011diffusion}, \textit{i.e.}, $Q z_0^2/D \approx 2.5396...$. The black solid line separates the different entropic regimes, \textit{i.e.}, values above, at and below the dashed line correspond to \textit{externally-driven} , \textit{zero-entropy} and \textit{Maxwell demon} resetting regimes, respectively. The intersection between both curves, marked by the star, occurs at $(c_0, 4\exp(-\gamma) c_0^2)$ where $c_0 \approx 1.90214$.} 
\label{fig:Qvsc}
\end{figure}

For $c\geq 1.8$ and as $c$ increases, the metastable minimum shifts to a lower $w$, 
while the MFPT value in the metastable minimum increases. 
Minimizing the MFPT with respect to $w$ at fixed $c$ leads to the transcedental equation:
\begin{equation}
    \frac{d\Theta(Q, \mathcal{N})}{dw}\frac{1}{1-\Theta(Q, \mathcal{N})} = \frac{2\Theta(Q, \mathcal{N})}{w} \; . \label{eq:trans}
\end{equation}
We have solved Eq. \eqref{eq:trans} numerically for different values of $c$ and in Fig. \ref{fig:Qvsc} we have plotted the minimizing resetting rate \textit{vs} $c$. We have added a horizontal  dashed line that corresponds to the resetting rate that minimizes the Evans-Majumdar stochastic resetting model. Notice that as $c$ increases, the MFPT minimum shifts to lower values of $Q$ eventually converging to $Q z_0^2/D \approx 2.5396...$ in perfect agreement with the Evans-Majumdar stochastic resetting model. Additionally, we have added a gray line that delimits the entropy regimes, \textit{viz.}, values above, at and below the dashed line correspond to  \textit{externally-driven}, \textit{zero-entropy} and \textit{Maxwell demon} regimes, respectively. We have colored the data points in order to show that, as $c$ increases and the local MFPT minimum shifts to the Maxwell demon regime, this MFPT value increases. We also computed the intersection between the locally optimal $Q$ and that defining the \textit{zero-entropy} regime denoted as a star in Fig. \ref{fig:Qvsc}. The intersection is located at $(c_0, 4w^2 c_0^2)$ where $c_0 \approx 1.90214$. In terms of the original variables, the MFPT metastable minimum corresponds to the \textit{zero-entropy} regime when $Q z_0^2/D \approx 4e^{-\gamma}(1.90214)^2 \approx 8.12575$.

As a matter of summary for this section, we have shown that in the case of a Brownian particle with stochastic reset to a random position sampled from a Gaussian distribution the MFPT is always finite. Furthermore, there is a metastable minimum at a finite resetting rate and as the Gaussian's standard deviation decreases, the MFPT metastable minimum eventually becomes the stable minimum. Additionally, we showed how the interplay between the resetting rate and the Gaussian standard deviation leads to three different entropy regimes. In the next section we study the effect of a toggling resetting distribution on the first passage properties.

\subsection{Stochastic resetting to random positions sampled from a toggling Gaussian distribution}
\label{sec:TogGauss}

In the previous section, we considered the resetting distributions to be Gaussian and showed that the MFPT metastable minimum can be tuned to lie in any of the thermodynamical regimes. In particular, for $w < \exp{(-\gamma/2)}$ the metastable minimum MFPT lies in the \textit{Maxwell demon} regime. Furthermore, we showed that the MFPT metastable minimum can match the \textit{zero-entropy} regime. Notice that ultimately the mean entropy per reset (Eq. \eqref{eq:EntPerRes}) is what allows us to define the three regimes. Here we consider the situation where a subset of the resetting events are such that the mean entropy per reset falls in the \textit{Maxwell demon} regime while for the remaining subset of resetting events the mean entropy per reset falls under the \textit{externally-driven} regime whereas the MFPE corresponds to the \textit{zero-entropy} regime.

Let us consider a Brownian particle that stochastically resets to a random position sampled from a \textit{toggling} Gaussian distribution. Depending on the resetting number parity, the particle resets to a random position sampled from $P_{a} = \mathcal{N}(\xi_i | z_{a}, \sigma_{a})$ or $P_{b} = \mathcal{N}(\xi_i | z_{b}, \sigma_{b})$ when the resetting number is even or odd, respectively. The resetting distribution set can be expressed as
\begin{equation}
    P_i(\xi_i) = \frac{1 + (-1)^{i}}{2} \mathcal{N}(\xi_i | z_{a}, \sigma_{a}) + \frac{1 + (-1)^{i+1}}{2} \mathcal{N}(\xi_i | z_{b}, \sigma_{b}) \; , \label{eq:PDF_TogglingGaussian}
\end{equation}
where $i$ tags the resetting number. From Eqs. \eqref{eq:Theta}, \eqref{eq:hallmarMFPT-3} and \eqref{eq:PDF_TogglingGaussian} we obtain
\begin{equation}
    \Theta(Q, P_{i}) = 
    \begin{cases}
        \Theta(Q, P_{a})  \; , & i = 2m \\
        \Theta(Q, P_{b})  \; , & i = 2m+1
    \end{cases} \label{eq:ThetaTogg} \; ,
\end{equation}
with $m$ an integer.
Let us introduce the same adimensional numbers and distinguish them by parity, \textit{i.e.},  $w_\chi = \sqrt{\sigma_\chi^2 Q/2D}$ and $c_\chi = z_\chi/\sqrt{2\sigma_\chi^2}$ with $\chi=\lbrace a,b\rbrace$.
Computing the MFPT and MFPNR is straighforward, \textit{viz.}, from Eq. \eqref{eq:ThetaTogg} and Eq. \eqref{eq:tau} we obtain the MFPT,
\begin{equation}
    \tau = \frac{\Theta(Q,P_{a})}{Q} \frac{1 + \Theta(Q,P_{b})}{1 - \Theta(Q,P_{b}) \cdot \Theta(Q,P_{a})} \; . \label{eq:twoTogglingPosG}
\end{equation}
The MFPNR can be directly inferred from Eq. \eqref{eq:twoTogglingPosG}. However, here we derive the expression from Eqs. \eqref{eq:Pnresets} and \eqref{eq:ThetaTogg}. Notice that the probability of finding the target after $n$ resets, $p_n$, is given by:
\begin{equation}
    p_n = \begin{cases}
        (1-\Theta(Q,P_{a})) (\Theta(Q,P_{a}) \Theta(Q,P_{b}))^m, \; \text{if } n=2m \\
        (1-\Theta(Q,P_{b})) (\Theta(Q,P_{a}) \Theta(Q,P_{b}))^m \Theta(Q,P_{a}), \; \\  \text{if } n=2m+1
    \end{cases} \; . 
\end{equation}
Thus the MFPNR (see Eq. \eqref{eq:mu}) yields:
\begin{equation}
    \mu = \Theta(Q,P_{a}) \frac{1 + \Theta(Q,P_{b})}{1 - \Theta(Q,P_{b}) \cdot \Theta(Q,P_{a})} \; . \label{eq:res2Gauss}
\end{equation}

Inferring the MFPE from the past results is not straightforward as the mean entropy per reset depends on the resetting number parity. With the aid of Eq. \eqref{eq:EntPerRes} we obtain the MFPE,
\begin{equation}
    \Delta \Sigma =  \frac{\left( \Delta s_Q(P_{b}) + \Theta(Q, P_{b}) \Delta s_Q(P_{a}) \right) \Theta(Q, P_{a})}{1-\Theta(Q, P_{b})\Theta(Q, P_{a})} \label{eq:Ent2Gauss}
\end{equation}

\begin{figure}[hbtp]
\centering
\includegraphics[width=3.1in]{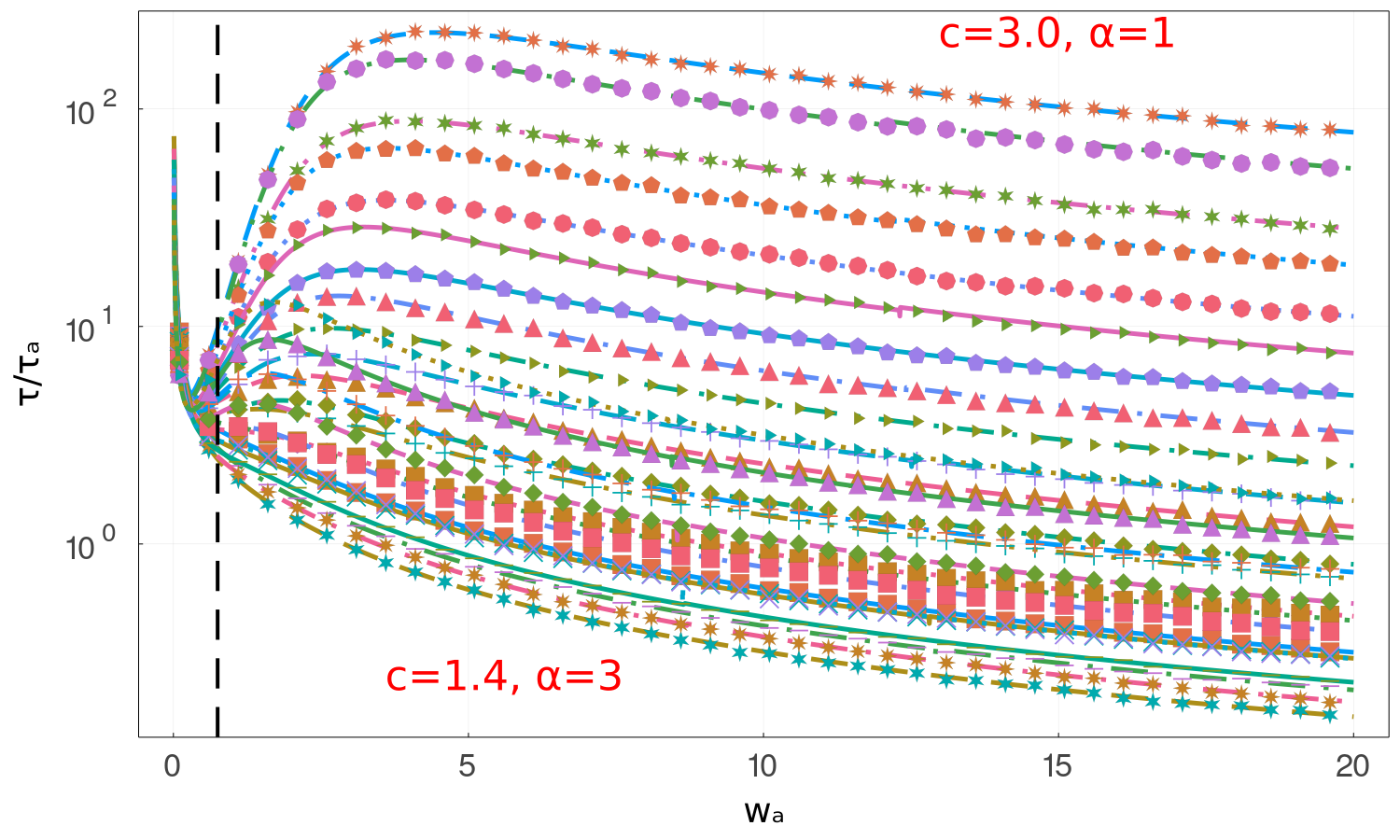}
\includegraphics[width=3.1in]{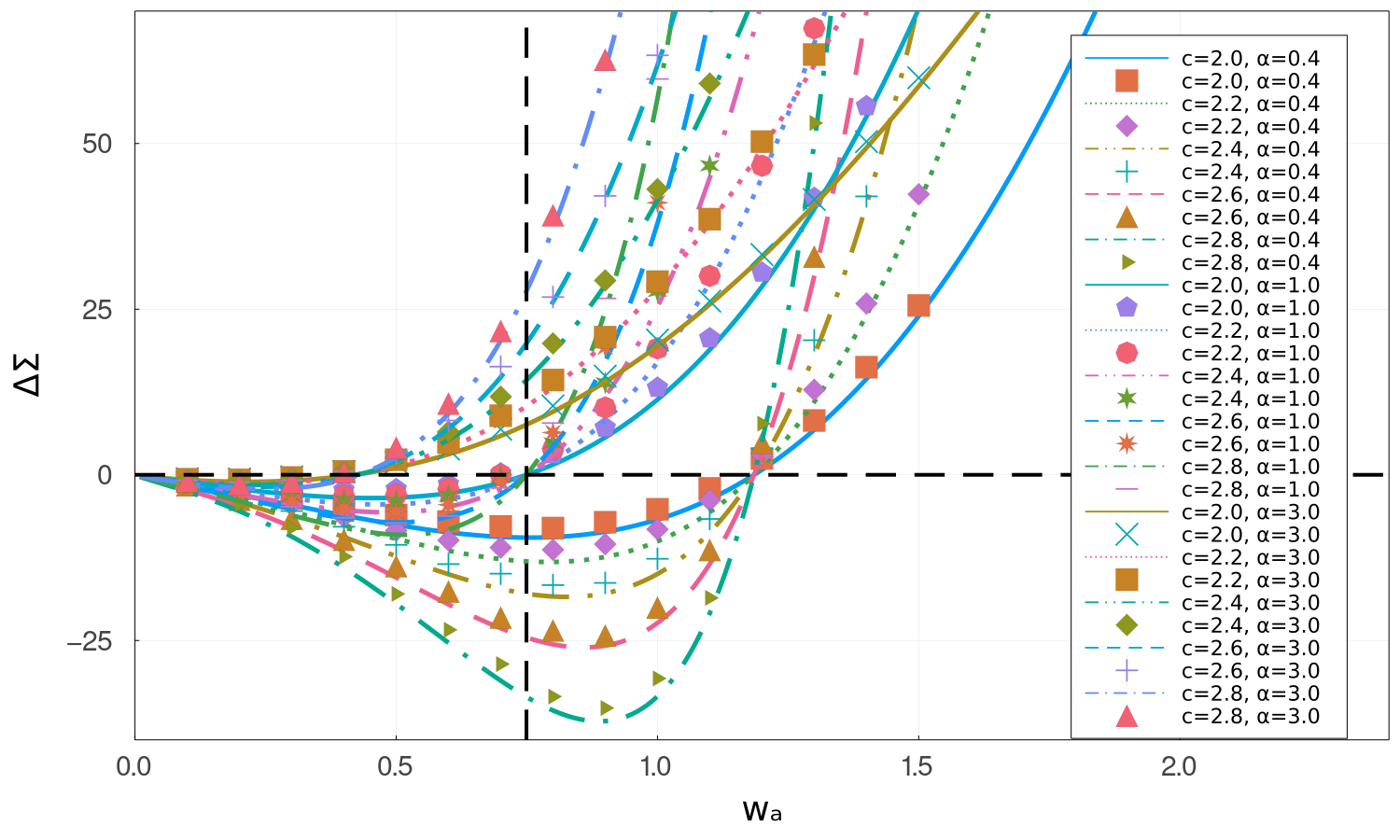}
\caption{Comparison between theory and simulations. Lines and plot markers correspond to theory and simulations respectively. The dashed vertical line delimits the entropy regimes per reset when $\alpha = 1$, while at the dashed horizontal corresponds to the \textit{zero-entropy} regime. \textbf{Upper panel} The reduced MFPT \textit{vs} $w$ for values of $(c_{a}, \alpha) \in [1.4,3.0)] \times [1.0,3.0]$ with steps $\Delta c = 0.2$ and $\Delta \alpha = 1.0$. For $c< 1.8$ the MFPT is decreasing monotonically. For $c\geq 1.8$, the MFPT has a metastable minimum and maximum close to the \textit{zero-entropy} regime. As $c$ increases to infinity, the metastable minimum becomes a stable minimum. \textbf{Lower panel} MFPE \textit{vs} $w$. The curves do not necessarily go through $(\exp(-\gamma/2), 0)$. As $c$ increases, the MFPE deviates more from zero for all $w$ values but $0$ and $\exp(-\gamma/2)$. The values of $z_a$ and $z_b$ were fixed to $0.1$ and $0.2$, respectively.  } \label{fig:2Gauss}
\end{figure}

Let us introduce the parameter $\alpha = \sigma_b/\sigma_a$, $w_b = \alpha w_a$ and $c_b \propto c_a/\alpha$. In Fig. \ref{fig:2Gauss} we compare Eqs. \eqref{eq:twoTogglingPosG} and \eqref{eq:Ent2Gauss} with simulations for different values of $c_a$ and $\alpha$ with $z_a=0.1$ and $z_b=0.2$. Notice in the upper panel that the MFPT also has a metastable minimum similar to that in the previous section. Furthermore, as $c$ increases, the metastable minimum shifts to a lower value of $w$. However, the entropy regime boundaries are no longer fixed as the MFPE depends on $w_a,c_a$ and $\alpha$. Additionally, while the comparison between theory and numerics in the lower panel is overall good, notice that for large $w_a$ and $c_a$, and hence larger resetting rate, the comparison improves.
When the typical resetting time is smaller than the typical time to diffuse to the origin the Brownian particle does not have enough time to probe the boundary and, therefore, $\mathcal{N}_{im} \approx \mathcal{N}$. In the case of small resetting rates the analytical results show a decent quantitative agreement and the same qualitative behavior as obtained numerically. We argue that the qualitative agreement comes from the fact that the information entropy associated to, both, the Gaussian distribution and to $\mathcal{N}_{im}(z | z_{n-1}, 2Dt)$ scales logarithmically with time.

Note that the MFPE belongs to the \textit{zero-entropy} regime when Eq. \eqref{eq:Ent2Gauss} equals zero, which leads to
\begin{equation}
    \Delta s_Q(P_{odd}) + \Theta(Q, P_{odd}) \Delta s_Q(P_{even}) = 0 \; , \label{eq:Ent2GaussSolve}
\end{equation}
while the entropy change per reset regime will depend on the resetting number parity. Thus the toggling feature allows a mechanism whereby the entropy gained per even resetting parity compensates for the entropy dissipated per odd resetting number (and \textit{vice versa}) leading to a net zero entropy first passage process. Furthermore, similar to the previous section, the MFPT metastable minimum can be in the \textit{zero-entropy} regime. Minimizing the MFPT with respect to $w_a$ leads to a transcendental equation which is reported in section \ref{sec:transcendentalEq} in the SM.

\begin{figure}[hbtp]
\centering
\includegraphics[width=3.1in]{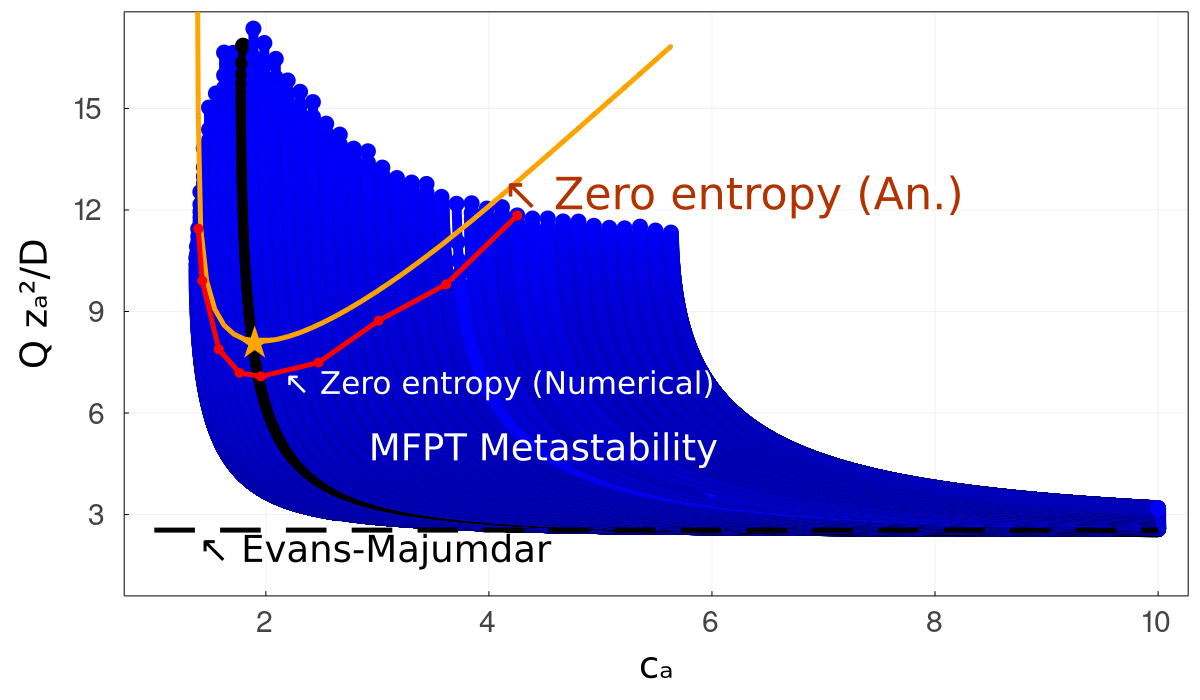}
\includegraphics[width=3.1in]{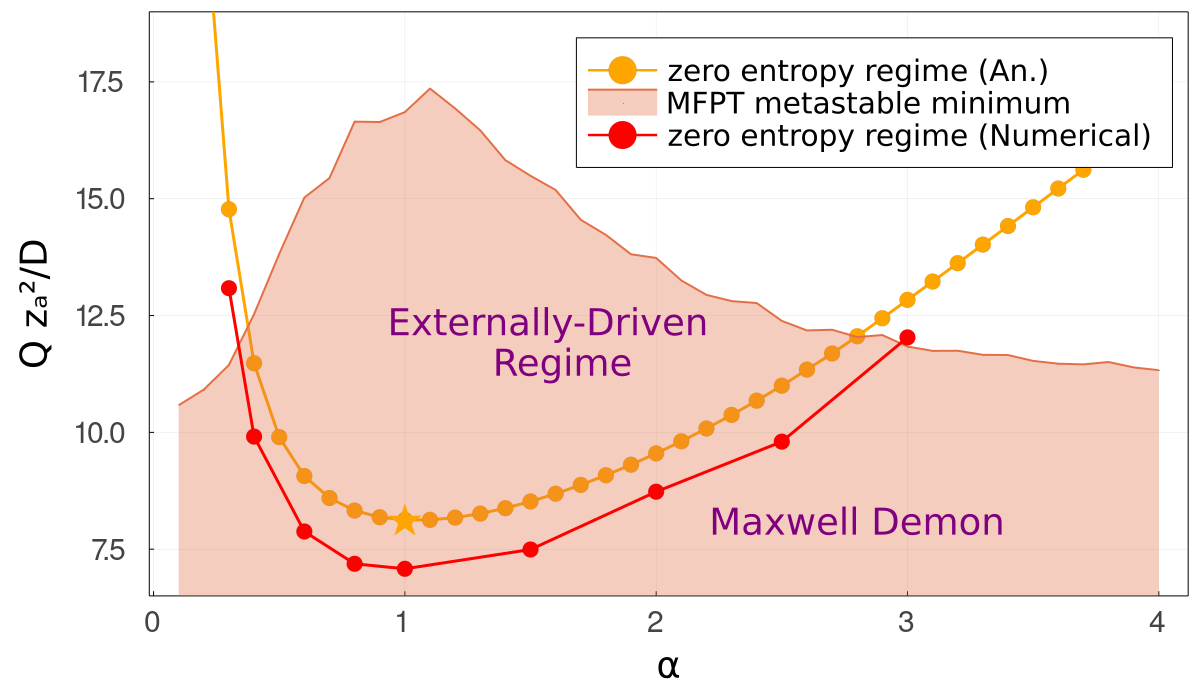}
\caption{MFPT metastable minimum obtained numerically from Eqs. \eqref{eq:Ent2GaussSolve}, \eqref{eq:MFPT2GaussSolve} and from numerical simulations. \textbf{Upper panel} Reduced resetting rate \textit{vs} $c_a$. Each blue curve correspond to a fix value of $\alpha$ where $\alpha \in [ 0.1, 4.0 ]$ with step $\Delta \alpha = 0.1$. The black curve correspond to $\alpha=1$ and larger values of $\alpha$ lie on the right hand side of the black curve. The dashed black horizontal line corresponds to the Evans-Majumdar limit where the MFPT minimum becomes absolute while the envelope of the blue curves limits the MFPT metastable minimum phase. The orange curve corresponds to the \textit{zero entropy} regime. Above and below are the \textit{externally-driven} and the \textit{Maxwell demon} regime. The yellow star corresponds to the MFPT metastable minimum falling in the \textit{zero entropy} regime for $\alpha=1 $. \textbf{Lower panel} Reduced resetting rate \textit{vs} $\alpha$. The filled band shows the metastable MFPT minimum whereas the yellow and red curves show the entropy regime.} \label{fig:TogglingGauss}
\end{figure}

\begin{figure}[hbtp]
\centering
\includegraphics[width=3.1in]{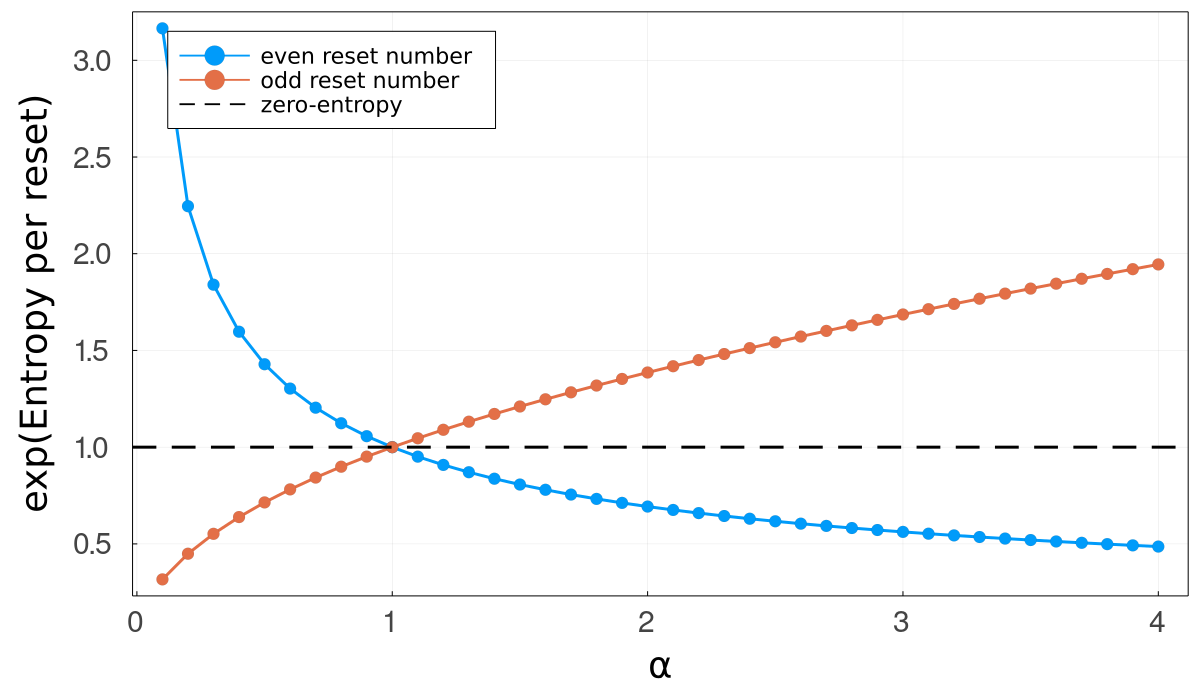}
\caption{Exponatiated entropy per reset \textit{vs} $\alpha$ for even and odd resets. The dashed horizontal line corresponds to the \textit{zero entropy} regime. Both curves are parametrized such that for any $\alpha$ value, the MFPE is zero.} \label{fig:TogglingGauss2}
\end{figure}

We solved Eqs. \eqref{eq:Ent2GaussSolve} and \eqref{eq:MFPT2GaussSolve} numerically for $w_a$ for different values of $c_a$ and $\alpha$. In the upper panel of Fig. \ref{fig:TogglingGauss} we show the numerical solutions by plotting the reduced resetting rate \textit{vs} $c_a$. Each blue curve corresponds to the MFPT mestastable minimum as a function of $c_a$ given a fixed value of $\alpha$, which increases in direction of the $x$-axis. Notice that the mestastable minimum disappears for low values of $c_a$, whereas for increasing $c_a$ the MFPT mestastable minimum becomes the stable minimum and converges to the Evans-Majumdar minimum. The solid orange line corresponds to the \textit{zero entropy} regime, above and below which are located the \textit{externally driven} and \textit{Maxwell demon} regimes, respectively.  Additionally, the red curve corresponds to the roots of the MFPE obtained from our numerical simulations which are equivalent to using the exact probability density function, \textit{i.e.} $\mathcal{N}_{im}(z|z_n, 2Dt)$. Note that there is a region approximately for $\alpha \in \left[ 0.4, 2.8\right] $ where the MFPT mestastable minimum matches the \textit{zero-entropy} regime as shown in the lower panel of Fig. \ref{fig:TogglingGauss}. Thus to any point in the red curve bounded by the blue curves corresponds a MFPT metastable minimum in the \textit{zero-entropy} regime. Moreover, for $\alpha \neq 0$, the entropy per reset for even (odd) resetting number falls in the \textit{Maxwell demon} (\textit{externally driven}) regime respectively, as shown in Fig. \ref{fig:TogglingGauss2}. Finally, our numerical results show that for $\alpha=1$ the MFPT metastable minimum corresponds to the \textit{zero-entropy} regime when $c_a \approx 1.95534$ and $Q z_a^2/D \approx 7.08033$.

To summarize, we have studied a Brownian particle with stochastic resetting to random positions sampled from a toggling distribution. Specifically, we computed the MFPT, MFPNR and the MFPE, and compared them with numerical simulations. We showed the existence of a region in parameter space where the MFPT metastable minimum belongs to the \textit{zero-entropy} regime while the entropy per reset can be non-zero. In this manner we were able to overlap the dynamic phase space with the entropy phase space.

\section{Conclusions}
\label{sec:Conclusions}
We have shown a novel method to compute the mean first passage time, number of resets and entropy for a one-dimensional Brownian particle subject to resetting to random positions, where at each reset the random position is sampled from a specific distribution associated to that resetting event. The method takes into account all of the resets happening in a first passage process and not just the last reset. This method is relevant since it allows to compute first passage statistics in the presence of stochastic resetting to random positions sampled from a reset-dependent distribution. We compared this method with simulations and found excellent agreement. In particular we first considered a generalized version of the Evans-Majumdar model and showed that the MFPT metastable minimum can be tuned to fall in either the \textit{energy-driven}, \textit{zero-entropy} or \textit{Maxwell demon} regime by tuning the resetting rate and the resetting distribution's first and second moment. We furthered considered the case of the parity-toggling resetting distribution and showed that it is possible to tune the parameters such that the MFPT metastable minimum corresponds to the \textit{zero-entropy} regime despite the average entropy per reset for even and odd resetting numbers is positive and negative, respectively. The toggling distribution ultimately shows how can stochastic resetting be utilized as a mechanism whereby information is randomly lost and gained \textit{locally} although the overall information flow is zero. Additionally, by studying the first passage time and entropy, we were able to compare the dynamic phase space with the entropic phase space.

Landauer's principle which relates the erasure of information to thermodynamic irreversibility states that an irreversible process in a computing device must necessarily come with an energy consumption. This principle has also been extended to biological settings used to model processes such as chemotaxis in single-cell organisms \cite{mehta2012energetic, wolpert2016free}. Understanding the dynamics and thermodynamics of search processes and how they are intertwined, can have applications in biology such as in RNA backtrack recovery \cite{dangkulwanich2013complete, roldan2016stochastic, ramoso2020stochastic}, in condensed matter for glass formation \cite{naumis2005energy, berthier2011theoretical, toledo2017short, toledo2018escape} and in deep learning for the training of deep neural networks\cite{toledo2021deep}. In this paper we have applied similar tools to resetting processes.


\section{Acknowledgements}
The JQTM acknowledges a Mitacs Accelerate Fellowship. The JQTM is thankful to Matteo Baggioli for reviewing the manuscript and providing useful comments.

\appendix
\renewcommand\appendixname{Supplemental Material}

\section{Numerical Simulations} \label{sec:Simulations}
In this section we describe the implementation our simulations. 
As in previous work \cite{toledo2019predator}, we performed a Kinetic Monte Carlo type simulation which first consists in generating a first passage time random variable, $t_0$, the L\'evy-Smirnov distribution with \textit{scale parameter} $z_0^2/2D$, as well as a resetting time random variable $t_{r0}$ from a Poisson distribution with parameter $Q$. If $t_0>t_{r0}$, we keep repeating the previous step. We denote the new first passage random 
variable as $t_n$ obtained from a L\'evy-Smirnov distribution, with scaling parameter $z_n^2/2D$ where $z_n$ is the random position after the resetting event, and denote the new resetting time random variable as $t_{rn}$, where $n$ 
denotes the number of resetting events that have occurred since the process started. If $t_n< t_{rn}$, then the process stops and the first passage time for that sample is 
$\sum_{i=0}^{n-1} t_{ri}+t_{n}$. Notice that $n$ is the first passage number of resets.
Similarly, the first passage entropy random variable is $\frac{1}{2} \sum_{i=0}^{n-1} \ln(\sigma^2_{i+1}/2Dt_{ri})$ where $\sigma_i$ is the $i$th resetting distribution's second moment. Our numerical results were obtained using a sample size of $5 \cdot 10^3$.

\begin{widetext}
\section{Mean First Passage time metastable minima} \label{sec:transcendentalEq}
We take the derivative w.r.t. $w_a$ of Eq. \eqref{eq:twoTogglingPosG}, equate to zero and obtain:
\begin{eqnarray}
   \left( \frac{d\Theta(Q, P_a)}{dw_a} +\frac{d\Theta(Q, P_a)}{dw_a} \Theta(Q, P_b) + \frac{d\Theta(Q, P_b)}{dw_a} \Theta(Q, P_b) \right)\frac{1}{\Theta(Q, P_a)(1+\Theta(Q, P_b)} + \nonumber \\
   \left(\frac{d\Theta(Q, P_a)}{dw_a} \Theta(Q, P_b) + \frac{d\Theta(Q, P_b)}{dw_a} \Theta(Q, P_a) \right) \frac{1}{1-\Theta(Q, P_a)\Theta(Q, P_b)}= \frac{2}{w_a} \; . \label{eq:MFPT2GaussSolve}
\end{eqnarray}
In Fig. \ref{fig:TogglingGauss} we display the root $w_a$ obtained numerically for fixed values of $c_a$ and $\alpha$.
\end{widetext}

\bibliography{references} 

\begin{thebibliography}{42}%
\makeatletter
\providecommand \@ifxundefined [1]{%
 \@ifx{#1\undefined}
}%
\providecommand \@ifnum [1]{%
 \ifnum #1\expandafter \@firstoftwo
 \else \expandafter \@secondoftwo
 \fi
}%
\providecommand \@ifx [1]{%
 \ifx #1\expandafter \@firstoftwo
 \else \expandafter \@secondoftwo
 \fi
}%
\providecommand \natexlab [1]{#1}%
\providecommand \enquote  [1]{``#1''}%
\providecommand \bibnamefont  [1]{#1}%
\providecommand \bibfnamefont [1]{#1}%
\providecommand \citenamefont [1]{#1}%
\providecommand \href@noop [0]{\@secondoftwo}%
\providecommand \href [0]{\begingroup \@sanitize@url \@href}%
\providecommand \@href[1]{\@@startlink{#1}\@@href}%
\providecommand \@@href[1]{\endgroup#1\@@endlink}%
\providecommand \@sanitize@url [0]{\catcode `\\12\catcode `\$12\catcode
  `\&12\catcode `\#12\catcode `\^12\catcode `\_12\catcode `\%12\relax}%
\providecommand \@@startlink[1]{}%
\providecommand \@@endlink[0]{}%
\providecommand \url  [0]{\begingroup\@sanitize@url \@url }%
\providecommand \@url [1]{\endgroup\@href {#1}{\urlprefix }}%
\providecommand \urlprefix  [0]{URL }%
\providecommand \Eprint [0]{\href }%
\providecommand \doibase [0]{https://doi.org/}%
\providecommand \selectlanguage [0]{\@gobble}%
\providecommand \bibinfo  [0]{\@secondoftwo}%
\providecommand \bibfield  [0]{\@secondoftwo}%
\providecommand \translation [1]{[#1]}%
\providecommand \BibitemOpen [0]{}%
\providecommand \bibitemStop [0]{}%
\providecommand \bibitemNoStop [0]{.\EOS\space}%
\providecommand \EOS [0]{\spacefactor3000\relax}%
\providecommand \BibitemShut  [1]{\csname bibitem#1\endcsname}%
\let\auto@bib@innerbib\@empty
\bibitem [{\citenamefont {Evans}\ and\ \citenamefont
  {Majumdar}(2011{\natexlab{a}})}]{evans2011diffusion}%
  \BibitemOpen
  \bibfield  {author} {\bibinfo {author} {\bibfnamefont {M.~R.}\ \bibnamefont
  {Evans}}\ and\ \bibinfo {author} {\bibfnamefont {S.~N.}\ \bibnamefont
  {Majumdar}},\ }\bibfield  {title} {\bibinfo {title} {Diffusion with
  stochastic resetting},\ }\href@noop {} {\bibfield  {journal} {\bibinfo
  {journal} {Physical review letters}\ }\textbf {\bibinfo {volume} {106}},\
  \bibinfo {pages} {160601} (\bibinfo {year} {2011}{\natexlab{a}})}\BibitemShut
  {NoStop}%
\bibitem [{\citenamefont {Evans}\ and\ \citenamefont
  {Majumdar}(2011{\natexlab{b}})}]{evans2011diffusionA}%
  \BibitemOpen
  \bibfield  {author} {\bibinfo {author} {\bibfnamefont {M.~R.}\ \bibnamefont
  {Evans}}\ and\ \bibinfo {author} {\bibfnamefont {S.~N.}\ \bibnamefont
  {Majumdar}},\ }\bibfield  {title} {\bibinfo {title} {Diffusion with optimal
  resetting},\ }\href@noop {} {\bibfield  {journal} {\bibinfo  {journal}
  {Journal of Physics A: Mathematical and Theoretical}\ }\textbf {\bibinfo
  {volume} {44}},\ \bibinfo {pages} {435001} (\bibinfo {year}
  {2011}{\natexlab{b}})}\BibitemShut {NoStop}%
\bibitem [{\citenamefont {Evans}\ \emph {et~al.}(2013)\citenamefont {Evans},
  \citenamefont {Majumdar},\ and\ \citenamefont {Mallick}}]{EvansJPhysA2013}%
  \BibitemOpen
  \bibfield  {author} {\bibinfo {author} {\bibfnamefont {M.~R.}\ \bibnamefont
  {Evans}}, \bibinfo {author} {\bibfnamefont {S.~N.}\ \bibnamefont
  {Majumdar}},\ and\ \bibinfo {author} {\bibfnamefont {K.}~\bibnamefont
  {Mallick}},\ }\bibfield  {title} {\bibinfo {title} {Optimal diffusive search:
  nonequilibrium resetting versus equilibrium dynamics},\ }\href
  {https://doi.org/10.1088/1751-8113/46/18/185001} {\bibfield  {journal}
  {\bibinfo  {journal} {Journal of Physics A: Mathematical and Theoretical}\
  }\textbf {\bibinfo {volume} {46}},\ \bibinfo {pages} {185001} (\bibinfo
  {year} {2013})}\BibitemShut {NoStop}%
\bibitem [{\citenamefont {Reuveni}(2016)}]{ReuveniPRL2016}%
  \BibitemOpen
  \bibfield  {author} {\bibinfo {author} {\bibfnamefont {S.}~\bibnamefont
  {Reuveni}},\ }\bibfield  {title} {\bibinfo {title} {Optimal stochastic
  restart renders fluctuations in first passage times universal},\ }\href
  {https://doi.org/10.1103/PhysRevLett.116.170601} {\bibfield  {journal}
  {\bibinfo  {journal} {Phys. Rev. Lett.}\ }\textbf {\bibinfo {volume} {116}},\
  \bibinfo {pages} {170601} (\bibinfo {year} {2016})}\BibitemShut {NoStop}%
\bibitem [{\citenamefont {Pal}\ \emph {et~al.}(2016)\citenamefont {Pal},
  \citenamefont {Kundu},\ and\ \citenamefont {Evans}}]{pal2016diffusion}%
  \BibitemOpen
  \bibfield  {author} {\bibinfo {author} {\bibfnamefont {A.}~\bibnamefont
  {Pal}}, \bibinfo {author} {\bibfnamefont {A.}~\bibnamefont {Kundu}},\ and\
  \bibinfo {author} {\bibfnamefont {M.~R.}\ \bibnamefont {Evans}},\ }\bibfield
  {title} {\bibinfo {title} {Diffusion under time-dependent resetting},\
  }\href@noop {} {\bibfield  {journal} {\bibinfo  {journal} {Journal of Physics
  A: Mathematical and Theoretical}\ }\textbf {\bibinfo {volume} {49}},\
  \bibinfo {pages} {225001} (\bibinfo {year} {2016})}\BibitemShut {NoStop}%
\bibitem [{\citenamefont {Pal}\ and\ \citenamefont
  {Reuveni}(2017)}]{pal2017first}%
  \BibitemOpen
  \bibfield  {author} {\bibinfo {author} {\bibfnamefont {A.}~\bibnamefont
  {Pal}}\ and\ \bibinfo {author} {\bibfnamefont {S.}~\bibnamefont {Reuveni}},\
  }\bibfield  {title} {\bibinfo {title} {First passage under restart},\
  }\href@noop {} {\bibfield  {journal} {\bibinfo  {journal} {Physical review
  letters}\ }\textbf {\bibinfo {volume} {118}},\ \bibinfo {pages} {030603}
  (\bibinfo {year} {2017})}\BibitemShut {NoStop}%
\bibitem [{\citenamefont {Toledo-Marin}\ \emph {et~al.}(2019)\citenamefont
  {Toledo-Marin}, \citenamefont {Boyer},\ and\ \citenamefont
  {Sevilla}}]{toledo2019predator}%
  \BibitemOpen
  \bibfield  {author} {\bibinfo {author} {\bibfnamefont {J.~Q.}\ \bibnamefont
  {Toledo-Marin}}, \bibinfo {author} {\bibfnamefont {D.}~\bibnamefont
  {Boyer}},\ and\ \bibinfo {author} {\bibfnamefont {F.~J.}\ \bibnamefont
  {Sevilla}},\ }\bibfield  {title} {\bibinfo {title} {Predator-prey dynamics:
  Chasing by stochastic resetting},\ }\href@noop {} {\bibfield  {journal}
  {\bibinfo  {journal} {arXiv preprint arXiv:1912.02141}\ } (\bibinfo {year}
  {2019})}\BibitemShut {NoStop}%
\bibitem [{\citenamefont {Gupta}(2019)}]{gupta2019stochastic}%
  \BibitemOpen
  \bibfield  {author} {\bibinfo {author} {\bibfnamefont {D.}~\bibnamefont
  {Gupta}},\ }\bibfield  {title} {\bibinfo {title} {Stochastic resetting in
  underdamped brownian motion},\ }\href@noop {} {\bibfield  {journal} {\bibinfo
   {journal} {Journal of Statistical Mechanics: Theory and Experiment}\
  }\textbf {\bibinfo {volume} {2019}},\ \bibinfo {pages} {033212} (\bibinfo
  {year} {2019})}\BibitemShut {NoStop}%
\bibitem [{\citenamefont {Evans}\ \emph {et~al.}(2020)\citenamefont {Evans},
  \citenamefont {Majumdar},\ and\ \citenamefont
  {Schehr}}]{evans2020stochastic}%
  \BibitemOpen
  \bibfield  {author} {\bibinfo {author} {\bibfnamefont {M.~R.}\ \bibnamefont
  {Evans}}, \bibinfo {author} {\bibfnamefont {S.~N.}\ \bibnamefont
  {Majumdar}},\ and\ \bibinfo {author} {\bibfnamefont {G.}~\bibnamefont
  {Schehr}},\ }\bibfield  {title} {\bibinfo {title} {Stochastic resetting and
  applications},\ }\href@noop {} {\bibfield  {journal} {\bibinfo  {journal}
  {Journal of Physics A: Mathematical and Theoretical}\ }\textbf {\bibinfo
  {volume} {53}},\ \bibinfo {pages} {193001} (\bibinfo {year}
  {2020})}\BibitemShut {NoStop}%
\bibitem [{\citenamefont {Tal-Friedman}\ \emph {et~al.}(2020)\citenamefont
  {Tal-Friedman}, \citenamefont {Pal}, \citenamefont {Sekhon}, \citenamefont
  {Reuveni},\ and\ \citenamefont {Roichman}}]{tal2020experimental}%
  \BibitemOpen
  \bibfield  {author} {\bibinfo {author} {\bibfnamefont {O.}~\bibnamefont
  {Tal-Friedman}}, \bibinfo {author} {\bibfnamefont {A.}~\bibnamefont {Pal}},
  \bibinfo {author} {\bibfnamefont {A.}~\bibnamefont {Sekhon}}, \bibinfo
  {author} {\bibfnamefont {S.}~\bibnamefont {Reuveni}},\ and\ \bibinfo {author}
  {\bibfnamefont {Y.}~\bibnamefont {Roichman}},\ }\bibfield  {title} {\bibinfo
  {title} {Experimental realization of diffusion with stochastic resetting},\
  }\href@noop {} {\bibfield  {journal} {\bibinfo  {journal} {The journal of
  physical chemistry letters}\ }\textbf {\bibinfo {volume} {11}},\ \bibinfo
  {pages} {7350} (\bibinfo {year} {2020})}\BibitemShut {NoStop}%
\bibitem [{\citenamefont {Mas{\'o}-Puigdellosas}\ \emph
  {et~al.}(2019)\citenamefont {Mas{\'o}-Puigdellosas}, \citenamefont {Campos},\
  and\ \citenamefont {M{\'e}ndez}}]{maso2019transport}%
  \BibitemOpen
  \bibfield  {author} {\bibinfo {author} {\bibfnamefont {A.}~\bibnamefont
  {Mas{\'o}-Puigdellosas}}, \bibinfo {author} {\bibfnamefont {D.}~\bibnamefont
  {Campos}},\ and\ \bibinfo {author} {\bibfnamefont {V.}~\bibnamefont
  {M{\'e}ndez}},\ }\bibfield  {title} {\bibinfo {title} {Transport properties
  of random walks under stochastic noninstantaneous resetting},\ }\href@noop {}
  {\bibfield  {journal} {\bibinfo  {journal} {Physical Review E}\ }\textbf
  {\bibinfo {volume} {100}},\ \bibinfo {pages} {042104} (\bibinfo {year}
  {2019})}\BibitemShut {NoStop}%
\bibitem [{\citenamefont {Krapivsky}\ and\ \citenamefont
  {Redner}(1996)}]{krapivsky1996kinetics}%
  \BibitemOpen
  \bibfield  {author} {\bibinfo {author} {\bibfnamefont {P.}~\bibnamefont
  {Krapivsky}}\ and\ \bibinfo {author} {\bibfnamefont {S.}~\bibnamefont
  {Redner}},\ }\bibfield  {title} {\bibinfo {title} {Kinetics of a diffusive
  capture process: lamb besieged by a pride of lions},\ }\href@noop {}
  {\bibfield  {journal} {\bibinfo  {journal} {J. Phys. A: Math. Gen.}\ }\textbf
  {\bibinfo {volume} {29}},\ \bibinfo {pages} {5347} (\bibinfo {year}
  {1996})}\BibitemShut {NoStop}%
\bibitem [{\citenamefont {Redner}\ and\ \citenamefont
  {Krapivsky}(1999)}]{redner1999capture}%
  \BibitemOpen
  \bibfield  {author} {\bibinfo {author} {\bibfnamefont {S.}~\bibnamefont
  {Redner}}\ and\ \bibinfo {author} {\bibfnamefont {P.}~\bibnamefont
  {Krapivsky}},\ }\bibfield  {title} {\bibinfo {title} {Capture of the lamb:
  Diffusing predators seeking a diffusing prey},\ }\href@noop {} {\bibfield
  {journal} {\bibinfo  {journal} {Am. J. Phys.}\ }\textbf {\bibinfo {volume}
  {67}},\ \bibinfo {pages} {1277} (\bibinfo {year} {1999})}\BibitemShut
  {NoStop}%
\bibitem [{\citenamefont {Oshanin}\ \emph {et~al.}(2009)\citenamefont
  {Oshanin}, \citenamefont {Vasilyev}, \citenamefont {Krapivsky},\ and\
  \citenamefont {Klafter}}]{oshanin2009survival}%
  \BibitemOpen
  \bibfield  {author} {\bibinfo {author} {\bibfnamefont {G.}~\bibnamefont
  {Oshanin}}, \bibinfo {author} {\bibfnamefont {O.}~\bibnamefont {Vasilyev}},
  \bibinfo {author} {\bibfnamefont {P.}~\bibnamefont {Krapivsky}},\ and\
  \bibinfo {author} {\bibfnamefont {J.}~\bibnamefont {Klafter}},\ }\bibfield
  {title} {\bibinfo {title} {Survival of an evasive prey},\ }\href@noop {}
  {\bibfield  {journal} {\bibinfo  {journal} {Proceedings of the National
  Academy of Sciences}\ }\textbf {\bibinfo {volume} {106}},\ \bibinfo {pages}
  {13696} (\bibinfo {year} {2009})}\BibitemShut {NoStop}%
\bibitem [{\citenamefont {Gabel}\ \emph {et~al.}(2012)\citenamefont {Gabel},
  \citenamefont {Majumdar}, \citenamefont {Panduranga},\ and\ \citenamefont
  {Redner}}]{gabel2012can}%
  \BibitemOpen
  \bibfield  {author} {\bibinfo {author} {\bibfnamefont {A.}~\bibnamefont
  {Gabel}}, \bibinfo {author} {\bibfnamefont {S.~N.}\ \bibnamefont {Majumdar}},
  \bibinfo {author} {\bibfnamefont {N.~K.}\ \bibnamefont {Panduranga}},\ and\
  \bibinfo {author} {\bibfnamefont {S.}~\bibnamefont {Redner}},\ }\bibfield
  {title} {\bibinfo {title} {Can a lamb reach a haven before being eaten by
  diffusing lions?},\ }\href@noop {} {\bibfield  {journal} {\bibinfo  {journal}
  {JSTAT}\ }\textbf {\bibinfo {volume} {2012}}\bibinfo  {number} { (05)},\
  \bibinfo {pages} {P05011}}\BibitemShut {NoStop}%
\bibitem [{\citenamefont {Redner}\ and\ \citenamefont
  {B{\'e}nichou}(2014)}]{redner2014gradual}%
  \BibitemOpen
\bibfield  {number} {  }\bibfield  {author} {\bibinfo {author} {\bibfnamefont
  {S.}~\bibnamefont {Redner}}\ and\ \bibinfo {author} {\bibfnamefont
  {O.}~\bibnamefont {B{\'e}nichou}},\ }\bibfield  {title} {\bibinfo {title}
  {Gradual diffusive capture: slow death by many mosquito bites},\ }\href@noop
  {} {\bibfield  {journal} {\bibinfo  {journal} {JSTAT}\ }\textbf {\bibinfo
  {volume} {2014}}\bibinfo  {number} { (11)},\ \bibinfo {pages}
  {P11019}}\BibitemShut {NoStop}%
\bibitem [{\citenamefont {Schwarzl}\ \emph {et~al.}(2016)\citenamefont
  {Schwarzl}, \citenamefont {Godec}, \citenamefont {Oshanin},\ and\
  \citenamefont {Metzler}}]{SchwarzlJPhysA2016}%
  \BibitemOpen
\bibfield  {number} {  }\bibfield  {author} {\bibinfo {author} {\bibfnamefont
  {M.}~\bibnamefont {Schwarzl}}, \bibinfo {author} {\bibfnamefont
  {A.}~\bibnamefont {Godec}}, \bibinfo {author} {\bibfnamefont
  {G.}~\bibnamefont {Oshanin}},\ and\ \bibinfo {author} {\bibfnamefont
  {R.}~\bibnamefont {Metzler}},\ }\bibfield  {title} {\bibinfo {title} {A
  single predator charging a herd of prey: effects of self volume and
  predator{\textendash}prey decision-making},\ }\href
  {https://doi.org/10.1088/1751-8113/49/22/225601} {\bibfield  {journal}
  {\bibinfo  {journal} {Journal of Physics A: Mathematical and Theoretical}\
  }\textbf {\bibinfo {volume} {49}},\ \bibinfo {pages} {225601} (\bibinfo
  {year} {2016})}\BibitemShut {NoStop}%
\bibitem [{\citenamefont {Das}\ and\ \citenamefont
  {Samanta}(2018)}]{DasJPhysA2018}%
  \BibitemOpen
  \bibfield  {author} {\bibinfo {author} {\bibfnamefont {A.}~\bibnamefont
  {Das}}\ and\ \bibinfo {author} {\bibfnamefont {G.~P.}\ \bibnamefont
  {Samanta}},\ }\bibfield  {title} {\bibinfo {title} {Modeling the fear effect
  on a stochastic prey{\textendash}predator system with additional food for the
  predator},\ }\href {https://doi.org/10.1088/1751-8121/aae4c6} {\bibfield
  {journal} {\bibinfo  {journal} {Journal of Physics A: Mathematical and
  Theoretical}\ }\textbf {\bibinfo {volume} {51}},\ \bibinfo {pages} {465601}
  (\bibinfo {year} {2018})}\BibitemShut {NoStop}%
\bibitem [{\citenamefont {Mercado-V{\'a}squez}\ and\ \citenamefont
  {Boyer}(2018)}]{mercado2018lotka}%
  \BibitemOpen
  \bibfield  {author} {\bibinfo {author} {\bibfnamefont {G.}~\bibnamefont
  {Mercado-V{\'a}squez}}\ and\ \bibinfo {author} {\bibfnamefont
  {D.}~\bibnamefont {Boyer}},\ }\bibfield  {title} {\bibinfo {title}
  {Lotka--volterra systems with stochastic resetting},\ }\href@noop {}
  {\bibfield  {journal} {\bibinfo  {journal} {Journal of Physics A:
  Mathematical and Theoretical}\ }\textbf {\bibinfo {volume} {51}},\ \bibinfo
  {pages} {405601} (\bibinfo {year} {2018})}\BibitemShut {NoStop}%
\bibitem [{\citenamefont {Boie}\ \emph {et~al.}(2018)\citenamefont {Boie},
  \citenamefont {Connor}, \citenamefont {McHugh}, \citenamefont {Nagel},
  \citenamefont {Ermentrout}, \citenamefont {Crimaldi},\ and\ \citenamefont
  {Victor}}]{boie2018information}%
  \BibitemOpen
  \bibfield  {author} {\bibinfo {author} {\bibfnamefont {S.~D.}\ \bibnamefont
  {Boie}}, \bibinfo {author} {\bibfnamefont {E.~G.}\ \bibnamefont {Connor}},
  \bibinfo {author} {\bibfnamefont {M.}~\bibnamefont {McHugh}}, \bibinfo
  {author} {\bibfnamefont {K.~I.}\ \bibnamefont {Nagel}}, \bibinfo {author}
  {\bibfnamefont {G.~B.}\ \bibnamefont {Ermentrout}}, \bibinfo {author}
  {\bibfnamefont {J.~P.}\ \bibnamefont {Crimaldi}},\ and\ \bibinfo {author}
  {\bibfnamefont {J.~D.}\ \bibnamefont {Victor}},\ }\bibfield  {title}
  {\bibinfo {title} {Information-theoretic analysis of realistic odor plumes:
  What cues are useful for determining location?},\ }\href@noop {} {\bibfield
  {journal} {\bibinfo  {journal} {PLoS computational biology}\ }\textbf
  {\bibinfo {volume} {14}},\ \bibinfo {pages} {e1006275} (\bibinfo {year}
  {2018})}\BibitemShut {NoStop}%
\bibitem [{\citenamefont {Baker}\ \emph {et~al.}(2018)\citenamefont {Baker},
  \citenamefont {Dickinson}, \citenamefont {Findley}, \citenamefont {Gire},
  \citenamefont {Louis}, \citenamefont {Suver}, \citenamefont {Verhagen},
  \citenamefont {Nagel},\ and\ \citenamefont {Smear}}]{baker2018algorithms}%
  \BibitemOpen
  \bibfield  {author} {\bibinfo {author} {\bibfnamefont {K.~L.}\ \bibnamefont
  {Baker}}, \bibinfo {author} {\bibfnamefont {M.}~\bibnamefont {Dickinson}},
  \bibinfo {author} {\bibfnamefont {T.~M.}\ \bibnamefont {Findley}}, \bibinfo
  {author} {\bibfnamefont {D.~H.}\ \bibnamefont {Gire}}, \bibinfo {author}
  {\bibfnamefont {M.}~\bibnamefont {Louis}}, \bibinfo {author} {\bibfnamefont
  {M.~P.}\ \bibnamefont {Suver}}, \bibinfo {author} {\bibfnamefont {J.~V.}\
  \bibnamefont {Verhagen}}, \bibinfo {author} {\bibfnamefont {K.~I.}\
  \bibnamefont {Nagel}},\ and\ \bibinfo {author} {\bibfnamefont {M.~C.}\
  \bibnamefont {Smear}},\ }\bibfield  {title} {\bibinfo {title} {Algorithms for
  olfactory search across species},\ }\href@noop {} {\bibfield  {journal}
  {\bibinfo  {journal} {Journal of Neuroscience}\ }\textbf {\bibinfo {volume}
  {38}},\ \bibinfo {pages} {9383} (\bibinfo {year} {2018})}\BibitemShut
  {NoStop}%
\bibitem [{\citenamefont {Dangkulwanich}\ \emph {et~al.}(2013)\citenamefont
  {Dangkulwanich}, \citenamefont {Ishibashi}, \citenamefont {Liu},
  \citenamefont {Kireeva}, \citenamefont {Lubkowska}, \citenamefont {Kashlev},\
  and\ \citenamefont {Bustamante}}]{dangkulwanich2013complete}%
  \BibitemOpen
  \bibfield  {author} {\bibinfo {author} {\bibfnamefont {M.}~\bibnamefont
  {Dangkulwanich}}, \bibinfo {author} {\bibfnamefont {T.}~\bibnamefont
  {Ishibashi}}, \bibinfo {author} {\bibfnamefont {S.}~\bibnamefont {Liu}},
  \bibinfo {author} {\bibfnamefont {M.~L.}\ \bibnamefont {Kireeva}}, \bibinfo
  {author} {\bibfnamefont {L.}~\bibnamefont {Lubkowska}}, \bibinfo {author}
  {\bibfnamefont {M.}~\bibnamefont {Kashlev}},\ and\ \bibinfo {author}
  {\bibfnamefont {C.~J.}\ \bibnamefont {Bustamante}},\ }\bibfield  {title}
  {\bibinfo {title} {Complete dissection of transcription elongation reveals
  slow translocation of rna polymerase ii in a linear ratchet mechanism},\
  }\href@noop {} {\bibfield  {journal} {\bibinfo  {journal} {Elife}\ }\textbf
  {\bibinfo {volume} {2}},\ \bibinfo {pages} {e00971} (\bibinfo {year}
  {2013})}\BibitemShut {NoStop}%
\bibitem [{\citenamefont {Rold{\'a}n}\ \emph {et~al.}(2016)\citenamefont
  {Rold{\'a}n}, \citenamefont {Lisica}, \citenamefont {S{\'a}nchez-Taltavull},\
  and\ \citenamefont {Grill}}]{roldan2016stochastic}%
  \BibitemOpen
  \bibfield  {author} {\bibinfo {author} {\bibfnamefont {{\'E}.}~\bibnamefont
  {Rold{\'a}n}}, \bibinfo {author} {\bibfnamefont {A.}~\bibnamefont {Lisica}},
  \bibinfo {author} {\bibfnamefont {D.}~\bibnamefont {S{\'a}nchez-Taltavull}},\
  and\ \bibinfo {author} {\bibfnamefont {S.~W.}\ \bibnamefont {Grill}},\
  }\bibfield  {title} {\bibinfo {title} {Stochastic resetting in backtrack
  recovery by rna polymerases},\ }\href@noop {} {\bibfield  {journal} {\bibinfo
   {journal} {Physical Review E}\ }\textbf {\bibinfo {volume} {93}},\ \bibinfo
  {pages} {062411} (\bibinfo {year} {2016})}\BibitemShut {NoStop}%
\bibitem [{\citenamefont {Zhang}\ and\ \citenamefont
  {Dudko}(2016)}]{zhang2016first}%
  \BibitemOpen
  \bibfield  {author} {\bibinfo {author} {\bibfnamefont {Y.}~\bibnamefont
  {Zhang}}\ and\ \bibinfo {author} {\bibfnamefont {O.~K.}\ \bibnamefont
  {Dudko}},\ }\bibfield  {title} {\bibinfo {title} {First-passage processes in
  the genome},\ }\href@noop {} {\bibfield  {journal} {\bibinfo  {journal}
  {Annual review of biophysics}\ }\textbf {\bibinfo {volume} {45}},\ \bibinfo
  {pages} {117} (\bibinfo {year} {2016})}\BibitemShut {NoStop}%
\bibitem [{\citenamefont {Mehta}\ and\ \citenamefont
  {Schwab}(2012)}]{mehta2012energetic}%
  \BibitemOpen
  \bibfield  {author} {\bibinfo {author} {\bibfnamefont {P.}~\bibnamefont
  {Mehta}}\ and\ \bibinfo {author} {\bibfnamefont {D.~J.}\ \bibnamefont
  {Schwab}},\ }\bibfield  {title} {\bibinfo {title} {Energetic costs of
  cellular computation},\ }\href@noop {} {\bibfield  {journal} {\bibinfo
  {journal} {Proceedings of the National Academy of Sciences}\ }\textbf
  {\bibinfo {volume} {109}},\ \bibinfo {pages} {17978} (\bibinfo {year}
  {2012})}\BibitemShut {NoStop}%
\bibitem [{\citenamefont {Wolpert}(2016)}]{wolpert2016free}%
  \BibitemOpen
  \bibfield  {author} {\bibinfo {author} {\bibfnamefont {D.~H.}\ \bibnamefont
  {Wolpert}},\ }\bibfield  {title} {\bibinfo {title} {The free energy
  requirements of biological organisms; implications for evolution},\
  }\href@noop {} {\bibfield  {journal} {\bibinfo  {journal} {Entropy}\ }\textbf
  {\bibinfo {volume} {18}},\ \bibinfo {pages} {138} (\bibinfo {year}
  {2016})}\BibitemShut {NoStop}%
\bibitem [{\citenamefont {Landauer}\ \emph {et~al.}(1991)\citenamefont
  {Landauer} \emph {et~al.}}]{landauer1991information}%
  \BibitemOpen
  \bibfield  {author} {\bibinfo {author} {\bibfnamefont {R.}~\bibnamefont
  {Landauer}} \emph {et~al.},\ }\bibfield  {title} {\bibinfo {title}
  {Information is physical},\ }\href@noop {} {\bibfield  {journal} {\bibinfo
  {journal} {Physics Today}\ }\textbf {\bibinfo {volume} {44}},\ \bibinfo
  {pages} {23} (\bibinfo {year} {1991})}\BibitemShut {NoStop}%
\bibitem [{\citenamefont {Blickle}\ \emph {et~al.}(2006)\citenamefont
  {Blickle}, \citenamefont {Speck}, \citenamefont {Helden}, \citenamefont
  {Seifert},\ and\ \citenamefont {Bechinger}}]{blickle2006thermodynamics}%
  \BibitemOpen
  \bibfield  {author} {\bibinfo {author} {\bibfnamefont {V.}~\bibnamefont
  {Blickle}}, \bibinfo {author} {\bibfnamefont {T.}~\bibnamefont {Speck}},
  \bibinfo {author} {\bibfnamefont {L.}~\bibnamefont {Helden}}, \bibinfo
  {author} {\bibfnamefont {U.}~\bibnamefont {Seifert}},\ and\ \bibinfo {author}
  {\bibfnamefont {C.}~\bibnamefont {Bechinger}},\ }\bibfield  {title} {\bibinfo
  {title} {Thermodynamics of a colloidal particle in a time-dependent
  nonharmonic potential},\ }\href@noop {} {\bibfield  {journal} {\bibinfo
  {journal} {Physical review letters}\ }\textbf {\bibinfo {volume} {96}},\
  \bibinfo {pages} {070603} (\bibinfo {year} {2006})}\BibitemShut {NoStop}%
\bibitem [{\citenamefont {B{\'e}rut}\ \emph {et~al.}(2012)\citenamefont
  {B{\'e}rut}, \citenamefont {Arakelyan}, \citenamefont {Petrosyan},
  \citenamefont {Ciliberto}, \citenamefont {Dillenschneider},\ and\
  \citenamefont {Lutz}}]{berut2012experimental}%
  \BibitemOpen
  \bibfield  {author} {\bibinfo {author} {\bibfnamefont {A.}~\bibnamefont
  {B{\'e}rut}}, \bibinfo {author} {\bibfnamefont {A.}~\bibnamefont
  {Arakelyan}}, \bibinfo {author} {\bibfnamefont {A.}~\bibnamefont
  {Petrosyan}}, \bibinfo {author} {\bibfnamefont {S.}~\bibnamefont
  {Ciliberto}}, \bibinfo {author} {\bibfnamefont {R.}~\bibnamefont
  {Dillenschneider}},\ and\ \bibinfo {author} {\bibfnamefont {E.}~\bibnamefont
  {Lutz}},\ }\bibfield  {title} {\bibinfo {title} {Experimental verification of
  landauer’s principle linking information and thermodynamics},\ }\href@noop
  {} {\bibfield  {journal} {\bibinfo  {journal} {Nature}\ }\textbf {\bibinfo
  {volume} {483}},\ \bibinfo {pages} {187} (\bibinfo {year}
  {2012})}\BibitemShut {NoStop}%
\bibitem [{\citenamefont {Jun}\ \emph {et~al.}(2014)\citenamefont {Jun},
  \citenamefont {Gavrilov},\ and\ \citenamefont {Bechhoefer}}]{jun2014high}%
  \BibitemOpen
  \bibfield  {author} {\bibinfo {author} {\bibfnamefont {Y.}~\bibnamefont
  {Jun}}, \bibinfo {author} {\bibfnamefont {M.}~\bibnamefont {Gavrilov}},\ and\
  \bibinfo {author} {\bibfnamefont {J.}~\bibnamefont {Bechhoefer}},\ }\bibfield
   {title} {\bibinfo {title} {High-precision test of landauer’s principle in
  a feedback trap},\ }\href@noop {} {\bibfield  {journal} {\bibinfo  {journal}
  {Physical review letters}\ }\textbf {\bibinfo {volume} {113}},\ \bibinfo
  {pages} {190601} (\bibinfo {year} {2014})}\BibitemShut {NoStop}%
\bibitem [{\citenamefont {MacKay}\ \emph {et~al.}(2003)\citenamefont {MacKay},
  \citenamefont {Mac~Kay} \emph {et~al.}}]{mackay2003information}%
  \BibitemOpen
  \bibfield  {author} {\bibinfo {author} {\bibfnamefont {D.~J.}\ \bibnamefont
  {MacKay}}, \bibinfo {author} {\bibfnamefont {D.~J.}\ \bibnamefont {Mac~Kay}},
  \emph {et~al.},\ }\href@noop {} {\emph {\bibinfo {title} {Information theory,
  inference and learning algorithms}}}\ (\bibinfo  {publisher} {Cambridge
  university press},\ \bibinfo {year} {2003})\BibitemShut {NoStop}%
\bibitem [{\citenamefont {Fuchs}\ \emph {et~al.}(2016)\citenamefont {Fuchs},
  \citenamefont {Goldt},\ and\ \citenamefont {Seifert}}]{fuchs2016stochastic}%
  \BibitemOpen
  \bibfield  {author} {\bibinfo {author} {\bibfnamefont {J.}~\bibnamefont
  {Fuchs}}, \bibinfo {author} {\bibfnamefont {S.}~\bibnamefont {Goldt}},\ and\
  \bibinfo {author} {\bibfnamefont {U.}~\bibnamefont {Seifert}},\ }\bibfield
  {title} {\bibinfo {title} {Stochastic thermodynamics of resetting},\
  }\href@noop {} {\bibfield  {journal} {\bibinfo  {journal} {EPL (Europhysics
  Letters)}\ }\textbf {\bibinfo {volume} {113}},\ \bibinfo {pages} {60009}
  (\bibinfo {year} {2016})}\BibitemShut {NoStop}%
\bibitem [{\citenamefont {Besga}\ \emph {et~al.}(2020)\citenamefont {Besga},
  \citenamefont {Bovon}, \citenamefont {Petrosyan}, \citenamefont {Majumdar},\
  and\ \citenamefont {Ciliberto}}]{besga2020optimal}%
  \BibitemOpen
  \bibfield  {author} {\bibinfo {author} {\bibfnamefont {B.}~\bibnamefont
  {Besga}}, \bibinfo {author} {\bibfnamefont {A.}~\bibnamefont {Bovon}},
  \bibinfo {author} {\bibfnamefont {A.}~\bibnamefont {Petrosyan}}, \bibinfo
  {author} {\bibfnamefont {S.~N.}\ \bibnamefont {Majumdar}},\ and\ \bibinfo
  {author} {\bibfnamefont {S.}~\bibnamefont {Ciliberto}},\ }\bibfield  {title}
  {\bibinfo {title} {Optimal mean first-passage time for a brownian searcher
  subjected to resetting: experimental and theoretical results},\ }\href@noop
  {} {\bibfield  {journal} {\bibinfo  {journal} {Physical Review Research}\
  }\textbf {\bibinfo {volume} {2}},\ \bibinfo {pages} {032029} (\bibinfo {year}
  {2020})}\BibitemShut {NoStop}%
\bibitem [{\citenamefont {Toledo-Marin}(2022)}]{githubJaque}%
  \BibitemOpen
  \bibfield  {author} {\bibinfo {author} {\bibfnamefont {J.~Q.}\ \bibnamefont
  {Toledo-Marin}},\ }\href@noop {} {\bibinfo {title} {First passage dynamics
  and information of a brownian particle with stochastic reset.}},\ \bibinfo
  {howpublished} {\url{https://github.com/jquetzalcoatl/BM-StochasticReset}}
  (\bibinfo {year} {2022})\BibitemShut {NoStop}%
\bibitem [{\citenamefont {Zwanzig}(2001)}]{zwanzig2001nonequilibrium}%
  \BibitemOpen
  \bibfield  {author} {\bibinfo {author} {\bibfnamefont {R.}~\bibnamefont
  {Zwanzig}},\ }\href@noop {} {\emph {\bibinfo {title} {Nonequilibrium
  statistical mechanics}}}\ (\bibinfo  {publisher} {Oxford University Press,
  USA},\ \bibinfo {year} {2001})\BibitemShut {NoStop}%
\bibitem [{\citenamefont {Evans}\ and\ \citenamefont
  {Majumdar}(2018)}]{evans2018run}%
  \BibitemOpen
  \bibfield  {author} {\bibinfo {author} {\bibfnamefont {M.~R.}\ \bibnamefont
  {Evans}}\ and\ \bibinfo {author} {\bibfnamefont {S.~N.}\ \bibnamefont
  {Majumdar}},\ }\bibfield  {title} {\bibinfo {title} {Run and tumble particle
  under resetting: a renewal approach},\ }\href@noop {} {\bibfield  {journal}
  {\bibinfo  {journal} {Journal of Physics A: Mathematical and Theoretical}\
  }\textbf {\bibinfo {volume} {51}},\ \bibinfo {pages} {475003} (\bibinfo
  {year} {2018})}\BibitemShut {NoStop}%
\bibitem [{\citenamefont {Ramoso}\ \emph {et~al.}(2020)\citenamefont {Ramoso},
  \citenamefont {Magalang}, \citenamefont {S{\'a}nchez-Taltavull},
  \citenamefont {Esguerra},\ and\ \citenamefont
  {Rold{\'a}n}}]{ramoso2020stochastic}%
  \BibitemOpen
  \bibfield  {author} {\bibinfo {author} {\bibfnamefont {A.~M.}\ \bibnamefont
  {Ramoso}}, \bibinfo {author} {\bibfnamefont {J.~A.}\ \bibnamefont
  {Magalang}}, \bibinfo {author} {\bibfnamefont {D.}~\bibnamefont
  {S{\'a}nchez-Taltavull}}, \bibinfo {author} {\bibfnamefont {J.~P.}\
  \bibnamefont {Esguerra}},\ and\ \bibinfo {author} {\bibfnamefont
  {{\'E}.}~\bibnamefont {Rold{\'a}n}},\ }\bibfield  {title} {\bibinfo {title}
  {Stochastic resetting antiviral therapies prevent drug resistance
  development},\ }\href@noop {} {\bibfield  {journal} {\bibinfo  {journal}
  {Europhysics Letters}\ }\textbf {\bibinfo {volume} {132}},\ \bibinfo {pages}
  {50003} (\bibinfo {year} {2020})}\BibitemShut {NoStop}%
\bibitem [{\citenamefont {Naumis}(2005)}]{naumis2005energy}%
  \BibitemOpen
  \bibfield  {author} {\bibinfo {author} {\bibfnamefont {G.~G.}\ \bibnamefont
  {Naumis}},\ }\bibfield  {title} {\bibinfo {title} {Energy landscape and
  rigidity},\ }\href@noop {} {\bibfield  {journal} {\bibinfo  {journal}
  {Physical Review E}\ }\textbf {\bibinfo {volume} {71}},\ \bibinfo {pages}
  {026114} (\bibinfo {year} {2005})}\BibitemShut {NoStop}%
\bibitem [{\citenamefont {Berthier}\ and\ \citenamefont
  {Biroli}(2011)}]{berthier2011theoretical}%
  \BibitemOpen
  \bibfield  {author} {\bibinfo {author} {\bibfnamefont {L.}~\bibnamefont
  {Berthier}}\ and\ \bibinfo {author} {\bibfnamefont {G.}~\bibnamefont
  {Biroli}},\ }\bibfield  {title} {\bibinfo {title} {Theoretical perspective on
  the glass transition and amorphous materials},\ }\href@noop {} {\bibfield
  {journal} {\bibinfo  {journal} {Reviews of modern physics}\ }\textbf
  {\bibinfo {volume} {83}},\ \bibinfo {pages} {587} (\bibinfo {year}
  {2011})}\BibitemShut {NoStop}%
\bibitem [{\citenamefont {Toledo-Mar{\'\i}n}\ and\ \citenamefont
  {Naumis}(2017)}]{toledo2017short}%
  \BibitemOpen
  \bibfield  {author} {\bibinfo {author} {\bibfnamefont {J.~Q.}\ \bibnamefont
  {Toledo-Mar{\'\i}n}}\ and\ \bibinfo {author} {\bibfnamefont {G.~G.}\
  \bibnamefont {Naumis}},\ }\bibfield  {title} {\bibinfo {title} {Short time
  dynamics determine glass forming ability in a glass transition two-level
  model: a stochastic approach using kramers’ escape formula},\ }\href@noop
  {} {\bibfield  {journal} {\bibinfo  {journal} {The Journal of Chemical
  Physics}\ }\textbf {\bibinfo {volume} {146}},\ \bibinfo {pages} {094506}
  (\bibinfo {year} {2017})}\BibitemShut {NoStop}%
\bibitem [{\citenamefont {Toledo-Mar{\'\i}n}\ and\ \citenamefont
  {Naumis}(2018)}]{toledo2018escape}%
  \BibitemOpen
  \bibfield  {author} {\bibinfo {author} {\bibfnamefont {J.~Q.}\ \bibnamefont
  {Toledo-Mar{\'\i}n}}\ and\ \bibinfo {author} {\bibfnamefont {G.~G.}\
  \bibnamefont {Naumis}},\ }\bibfield  {title} {\bibinfo {title} {Escape time,
  relaxation, and sticky states of a softened henon-heiles model: Low-frequency
  vibrational mode effects and glass relaxation},\ }\href@noop {} {\bibfield
  {journal} {\bibinfo  {journal} {Physical Review E}\ }\textbf {\bibinfo
  {volume} {97}},\ \bibinfo {pages} {042106} (\bibinfo {year}
  {2018})}\BibitemShut {NoStop}%
\bibitem [{\citenamefont {Toledo-Mar{\'\i}n}\ \emph {et~al.}(2021)\citenamefont
  {Toledo-Mar{\'\i}n}, \citenamefont {Fox}, \citenamefont {Sluka},\ and\
  \citenamefont {Glazier}}]{toledo2021deep}%
  \BibitemOpen
  \bibfield  {author} {\bibinfo {author} {\bibfnamefont {J.~Q.}\ \bibnamefont
  {Toledo-Mar{\'\i}n}}, \bibinfo {author} {\bibfnamefont {G.}~\bibnamefont
  {Fox}}, \bibinfo {author} {\bibfnamefont {J.~P.}\ \bibnamefont {Sluka}},\
  and\ \bibinfo {author} {\bibfnamefont {J.~A.}\ \bibnamefont {Glazier}},\
  }\bibfield  {title} {\bibinfo {title} {Deep learning approaches to surrogates
  for solving the diffusion equation for mechanistic real-world simulations},\
  }\href@noop {} {\bibfield  {journal} {\bibinfo  {journal} {Frontiers in
  Physiology}\ }\textbf {\bibinfo {volume} {12}} (\bibinfo {year}
  {2021})}\BibitemShut {NoStop}%
\end{thebibliography}%


\end{document}